\documentclass[12pt]{article}
\pdfoutput=1
\usepackage{subfigure}
\usepackage{amssymb,amsmath}
\usepackage{graphicx}
\usepackage{color}
\usepackage[colorlinks=true
,urlcolor=blue
,citecolor=blue
,linkcolor=blue
,pagecolor=blue
,linktocpage=true
,pdfproducer=medialab
]{hyperref}
\usepackage[a4paper,width=15.2cm]{geometry}
\usepackage{multirow}

\makeatletter \renewcommand{\@dotsep}{10000} \makeatother

\def\tmu{\tilde \mu}

\def\tw{\widetilde \chi^{\pm}}

\newcommand{\beq}{\begin{equation}}
\newcommand{\eeq}{\end{equation}}
\newcommand{\bea}{\begin{eqnarray}}
\newcommand{\eea}{\end{eqnarray}}

\newcommand{\neu}[1]{\ensuremath{\tilde{\chi}_{#1}^0}}
\newcommand{\chp}[1]{\ensuremath{\tilde{\chi}_{#1}^+}}

\newcommand{\chpm}[1]{\ensuremath{\tilde{\chi}_{#1}^{\pm}}}
\newcommand{\chmp}[1]{\ensuremath{\tilde{\chi}_{#1}^{\mp}}}

\newcommand{\gmu}{\ensuremath{(g-2)_{\mu}}}

\newcommand{\ttbar}{t \bar{t}}

\newcommand{\met} {{E\!\!\!\!/_{T}}}

\newcommand{ \pythia } {{\tt PYTHIAv6}}

\newcommand{ \pgs }    {{\tt PGS4}}

\newcommand{ \madgraph } {{\tt MADGRAPH5}}
\newcommand{ \micromegas } {{\tt micrOMEGAs4.1}}
\newcommand{ \FH } {{\tt FeynHiggs}}
\newcommand{ \RNORMX } {{\tt RNORMX}}
\newcommand{\CheckMATE}{CheckMATE}

\def\gev{\,{\rm GeV}}
\def\tev{\,{\rm TeV}}

\begin{document}

\begin{flushright}
MI-TH-1514
\end{flushright}

\begin{center}

 {\Large\bf  Neutralinos and Sleptons at the LHC in Light of Muon $\gmu$ 
 } \vspace{1cm}

{  M. Adeel Ajaib$^{a,}$\footnote{ E-mail: adeel@udel.edu}, Bhaskar Dutta$^{b,}$\footnote{ E-mail: dutta@physics.tamu.edu}, Tathagata Ghosh$^{b,}$\footnote{ E-mail: ghoshtatha@physics.tamu.edu},\\ Ilia Gogoladze$^{c,}$\footnote{E-mail:
ilia@bartol.udel.edu\\ \hspace*{0.5cm} On  leave of absence from:
Andronikashvili Institute of Physics, 0177 Tbilisi, Georgia.},  
  Qaisar Shafi$^{c,}$\footnote{E-mail:
shafi@bartol.udel.edu} } \vspace{.5cm}

{ \it 
$^a$ Department of Physics and Astronomy, Ursinus College, \\ Collegeville, PA 19426, USA \\

\vspace*{3mm}
$^b$Mitchell Institute of Fundamental Physics and Astronomy, Department of Physics and Astronomy,
Texas A\&M University, College Station, TX 77843-4242, USA  \\

\vspace*{3mm}
$^c$Bartol Research Institute, Department of Physics and Astronomy, \\
University of Delaware, Newark, DE 19716, USA 

} \vspace{.5cm}

\vspace{1.5cm}
 {\bf Abstract}\end{center}

  We study the neutralinos and sleptons in multi-lepton final states at the LHC in light of $\gmu$ anomaly. We scan the MSSM parameters relevant to $\gmu$  and focus on three distinct cases with different neutralino compositions. The explanation  of $\gmu$ excess at 2$\sigma$ range requires the smuon ($\tilde{\mu}_1$) to be lighter than $\sim$ 500 (1000) GeV for $\tan \beta=10\,(50)$. Correspondingly the two lightest neutralinos, $\neu{1},\neu{2}$, have to be lighter than $\sim$ 300 (650) GeV and 900 (1500) GeV respectively. We explore the prospects of searching these  light neutralinos and smuons at the LHC.
 The upcoming run of the LHC will be able to set $95\%$ CL exclusion limit on $M_{\neu{2}}$ ($\sim 650 - 1300$ GeV) and $m_{\tilde{l}}$ ($\sim 670-775$ GeV) with $M_{\neu{1}} \sim 100-250$ GeV at 3000 fb$^{-1}$ integrated luminosity in multi-lepton + $\met$ channel. 

\newpage

\renewcommand{\thefootnote}{\arabic{footnote}}
\setcounter{footnote}{0}

\section{Introduction}

  The LHC experiments have been a resounding success so far with the discovery of a Standard Model (SM)-like Higgs boson \cite{:2012gk, :2012gu}, but any signature of physics beyond the SM remaining elusive. The observed Higgs boson mass by CMS and ATLAS has strengthened the argument for weak-scale Supersymmetry (SUSY),  since the Minimal Supersymmetric Standard Model (MSSM) predicts an upper bound of $m_h \lesssim 135$ GeV for the lightest  CP-even Higgs boson  \cite{Carena:2002es}. However to definitively prove the  weak-scale realization of SUSY in nature, the discovery of supersymmetric partners of the SM electroweak (EW) particles is of paramount importance. Within the framework of MSSM the lightest neutralino ($\neu{1}$), is a compelling Dark Matter(DM) candidate, which constitutes nearly 80$\%$ of the matter in the Universe. Consequently it is of great significance to probe the EW sector of SUSY models, especially the composition of $\neu{1}$, to understand it's connection to the DM. 
 
  It is also well known that weak-scale SUSY can accommodate the $2-3 \, \sigma$ discrepancy between the measurement of $\gmu$ by the BNL~\cite{BNL} experiment and its value predicted by the SM. It requires the existence of relatively light smuon and gaugino (wino or bino).
BNL has measured an excess of $\sim 3.6 \, \sigma \, (2.4 \, \sigma)$ in $(g-2)_{\mu}$, using $e^+e^- \rightarrow$ hadrons (hadronically decaying $\tau$) data~\cite{BNL,Davier:2013wwg}. Various theoretical computations within the SM~\cite{Hagiwara:2011af,Benayoun:2012wc,Davier:2010nc} have been performed by different groups to explain this excess, but to no avail. The deviation in $(g-2)_{\mu}$ from the SM prediction is:
\begin{eqnarray}
\Delta a_{\mu}\equiv a_{\mu}({\rm exp})-a_{\mu}({\rm SM})= (28.6 \pm 8.0) \times 10^{-10}
\end{eqnarray}

In this paper we perform a weak-scale MSSM scan in order to study the parameter space that resolves the $(g-2)_{\mu}$ anomaly. There have been several recent attempts to resolve this discrepancy within the MSSM framework assuming non-universal SUSY-breaking (SSB) mass terms at $M_{\rm GUT}$ for gauginos~\cite{Akula:2013ioa,Pokoroski,Yokozaki} or  sfermions~\cite{Baer:2004xx,Ibe:2013oha}. The novel features of our analysis include highlighting the composition of the neutralinos that resolves the $(g-2)_{\mu}$ anomaly and the corresponding signal predictions at the upcoming 14 TeV run of the LHC. Previously Ref.~\cite{Endo} studied electroweakinos at 8 TeV using cascade decay of gluinos and in $3 l + \met$ channel, but for GUT constraint $M_2= 2 M_1$ only. In recent studies Refs.~\cite{Kuver,Kowalska:14TeV} have also investigated the prospect of $\gmu$ at LHC14. While Ref.~\cite{Kuver} has focussed on light stop assisted scenarios only, motivated by naturalness argument, Ref.~\cite{Kowalska:14TeV} has performed a broader study for GUT-constrained scenarios. However Ref.~\cite{Kowalska:14TeV} has derived their exclusion limits on electroweakino masses, without identifying their nature, based on kinematic cuts devised by CMS and ATLAS for 8 TeV in $2 \, l$ and $3 \, l$ final states. In contrast we systematically studied the contents of electroweakinos, model independently, without any a-priori high energy or fine-tuning conditions, and set exclusion limits using all possible multi-lepton channels. Furthermore we enriched the existing CMS and ATLAS search strategies with the inclusion of additional kinematic cuts. 

    The allowed parameter space of the MSSM will be heavily constrained if we require neuttralino LSP to satisfy observed DM relic density as well as constraints arising from indirect and direct DM detection searches. However the constraints from direct detection experiments suffer from large uncertainties in proton properties. 
Indirect detection constraints also suffer from uncertainties in various astrophysical factors. 
  Hence, in this paper, we did not restrict ourselves to relic density or DM direct and indirect searches allowed regions but 
  commented on them occasionally. However, if the DM constraints are applied, the reach for the SUSY particles pertaining to $\gmu$ parameter space, can easily be obtained from our results. 

The paper is organized as follows. In Section \ref{g-2} we briefly describe the expression for the SUSY contribution to $\gmu$ in the MSSM. In Section \ref{pheno} we summarize the scanning procedure and the general classifications of the parameter space. In Section \ref{results-1} we present the bounds on the relevant superpartner masses from \gmu \, and commented on possible DM constraints. The production mechanism of electroweakinos are discussed in Section \ref{LHC}, together with the simulation methods we adopted for this analysis. In Section \ref{results-2} we discuss the prospects of electroweakino and smuon searches in the present and upcoming runs of the LHC. In Section \ref{conclusions} we present our conclusions.

\section{\label{g-2}The Muon Anomalous Magnetic Moment}

The leading  contribution from low scale supersymmetry  to the muon anomalous magnetic moment is given by \cite{Moroi:1995yh, Martin:2001st}:

\bea
\label{eqq1}
\Delta a_\mu &=& \frac{\alpha \, m^2_\mu \, \mu\,  \tan\beta}{4\pi} {\bigg \{ }
\frac{M_{2}}{ \sin^2\theta_W \, m_{\tilde{\mu}_{L}}^2}
\left[ \frac{f_{\chi}(M_{2}^2/m_{\tilde{\mu}_{L}}^2)-f_{\chi}(\mu^2/m_{\tilde{\mu}_{L}}^2)}{M_2^2-\mu^2} \right] 
\nonumber\\
&+&
\frac{M_{1} }{ \cos^2\theta_W \, (m_{\tilde{\mu}_{R}}^2 - m_{\tilde{\mu}_{L}}^2)}
\left[\frac{f_{N}(M^2_1/m_{\tilde{\mu}_{R}}^2)}{m_{\tilde{\mu}_{R}}^2} - \frac{f_{N}(M^2_1/m_{\tilde{\mu}_{L}}^2)}{m_{\tilde{\mu}_{L}}^2}\right] \, {\bigg \} },
\eea
 where $\alpha$ is the fine-structure constant, $m_\mu$ is the muon mass, $\mu$ denotes  the bilinear Higgs mixing term, and $\tan\beta$ is the ratio of the vacuum expectation values (VEV) of the MSSM Higgs doublets. $M_1$ and $M_2$ denote the $U(1)_Y$ and $SU(2)$ gaugino masses respectively, $\theta_W$  is the weak mixing angle, and $m_{\tilde{\mu}_{L}}$ and $m_{\tilde{\mu}_{R}}$ are the left and right handed smuon masses. The loop functions are defined as follows:
\bea
f_{\chi}(x) &=& \frac{x^2 - 4x + 3 + 2\ln x}{(1-x)^3}~,\qquad ~f_{\chi}(1)=-2/3, \\
f_{N}(x) &=& \frac{ x^2 -1- 2x\ln x}{(1-x)^3}\,,\qquad\qquad f_{N}(1) = -1/3 \, .
\label{eqq2}
\eea
The first term in equation (\ref{eqq1}) stands for the dominant contribution coming from one loop diagram with charginos (Higgsinos and Winos), while the second term describes inputs from bino-smuon loop.
As the Higgsino mass $\mu$ increases, the first term decreases in equation (\ref{eqq1}), while the second term becomes dominant. On the other hand the smuons  needs to be light, $\mathcal{O}$(500 GeV), in both cases in order to make sizeable contribution to $\gmu$. Note that equation (\ref{eqq1}) will eventually fail to be accurate for a very big values of $\mu\tan\beta$, according to the decoupling theory.
As equation (\ref{eqq1}) indicates, the parameters
\begin{equation}
 M_1, \, M_2, \, \mu,\, \tan\beta,  m_{\tilde{\mu}_{L}}, \, m_{\tilde{\mu}_{R}},
\label{eqq3}
\end{equation}
are most relevant for the $\gmu$.

\section{Parameter Space and General Classification}\label{pheno}

  In this section we briefly discuss our scanning procedure and the method of classification of the parameter space subject to the composition of electroweakinos. As highlighted earlier, the BNL measured $\gmu$ differs from the SM prediction by more than $2 \sigma$. In this paper we employ the following $1\sigma$ and $2 \sigma$ ranges of $\gmu$:

\begin{align}
12.6 \times 10^{-10} <  \Delta a_\mu  < 44.6 \times 10^{-10} \,\, , && (2\sigma) \label{gm2-2s} \\
20.6 \times 10^{-10} <  \Delta a_\mu  < 36.6 \times 10^{-10} \,\, . && (1\sigma) \label{gm2-1s}
\end{align}
%

It has been noted in previous studies that  smuon and electroweakino masses upto $\sim 1$ TeV can resolve the $\gmu$ anomaly in various settings of MSSM~\cite{Akula:2013ioa,Pokoroski,Baer:2004xx,Ibe:2013oha}. This motivates us to search for these light smuons and electroweakinos at the upcoming high luminosity 14 TeV run of the LHC, in a model independent way. In its previous run, the LHC has set impressive bounds [$\mathcal{O}$(TeV] on squark and gluino masses. Although the squarks and gluinos have no direct influence on $\gmu$, they being heavy prohibits an abundant production of electroweakinos and smuons through cascade decays. We are thus restricted to probe electroweakinos and smuons by means of their direct production at the LHC. 

  We can study the SUSY particles, involved in $\gmu$ from three different directions. First, we can search for the neutralino LSP by adopting the monojet~\cite{CompressedGaugino,Han,Baer:2014kya,Han:2013usa,Cirelli:wino,Baer:Puremono} or vector boson fusion (VBF)~\cite{VBF-DM} search strategies. However, these searches will not yield any insight about the detailed particle spectrum needed to calculate $\gmu$. Moreover as shown above a vast amount of work has been done in the literature to detect neutalino LSP at the LHC. Hence we have not performed any rigorous analysis in this direction but extracted and extrapolated results from the references mentioned above.
The second and more promising approach is to search for heavier neutralinos and charginos. Searching for them are of particular importance when the LSP is bino-like due to extremely low production rate of bino at the LHC~\cite{VBF-DM}. The 14 TeV LHC will still produce these particles sufficiently, due to the presence of large wino and higgsino components in their compositions. We have looked for these heavier neutralinos and charginos in inclusive searches involving multilepton + $\met$ final states over a vast region of MSSM parameter space. Finally, one can search for smuons directly at the LHC but their production is also kinematically suppressed. Although we did not carry out any exclusive search for smuons, whenever necessary we have extrapolated the results from Ref.~\cite{Eckel:2014dza}, where the authors have explored the prospect of finding sleptons at the LHC for different compositions of the LSP.      

Having outlined our motivation for the paper let us discuss the scanning procedure of the parameter space in more detail. We employ the \FH~\cite{FeynHiggs} package to randomly scan the parameters relevant for SUSY contribution to $\gmu$. 
In performing the random scan a uniform and logarithmic distribution of random points is first generated in the selected parameter space.
The function \RNORMX \cite{Leva} is then employed
to generate a gaussian distribution around each point in the parameter space.   We set the top quark mass  $m_t = 173.3\, {\rm GeV}$  \cite{:1900yx}. The range of the parameters we scan are as follows:

\begin{align}
0 &<  M_1  <  1 \tev \, , \nonumber \\
0 &<  M_2  <  1 \tev  \, , \nonumber \\
0 &<  m_{\tilde{\mu}_L} <  1 \tev \, , \nonumber \\
0 &<   m_{\tilde{\mu}_R} <  1 \tev \, , \nonumber \\
0 &<   \mu <  1 \tev \, . 
\label{parameterRange}
\end{align}
%
 Here $M_1$, $M_2$ are the bino and wino SSB mass terms at the weak-scale, and  $m_{\tilde{\mu}_L}$ and $m_{\tilde{\mu}_R}$ are the left and right handed smuon SSB mass terms respectively. Two values of $\tan \beta$ - 10 and 50 have been chosen for the scanning procedure. All other mass parameters are set equal to 5 TeV and the A-terms were set equal to zero. We require degeneracy among the first and second generation slepton masses in order to be consistent with the constraints from $\mu \rightarrow e \gamma$ flavor-changing neutral current (FCNC) process. The dependence of $\gmu$ on the remaining SUSY mass parameters are negligible and they are kept heavy [$\mathcal{O} $(TeV)]. 

 The SUSY contribution to $\gmu$ is largest, if $M_1$, $M_2$ and $\mu$ have the same sign~\cite{Pokoroski}. In this case both terms in equation (\ref{eqq1}), arising from chargino-sneutrino and bino-smuon loops respectively, will be positive. Although we have limited our scan to  positive values of $M_1$, $M_2$ and $\mu$, and  $\gmu$ is satisfied when all of them have negative sign as well, but simultaneous change of sign will have no impact on the mass spectrum of the electroweakinos, which is the main ingredient of our collider analysis. Furthermore, we should point out that despite having scanned  $M_1$, $M_2$ and $\mu$ upto 1 TeV only for the plots presented in the paper, we have explored scenarios beyond 1 TeV  whenever the collider study required it.  

 In addition We apply the following LEP constraints~\cite{LEP} on the data that we acquire from \FH:

\begin{align}
m_{\tmu_{L,R}} > 100 \gev \, ,\nonumber \\
M_{\tw_1} > 105 \,    \gev. 
\label{mass-constraints}
\end{align}
We also impose the lower bound on the $\neu{1}$ mass, $M_{\neu{1}}>53 \gev$ if the $\neu{1}$ is not a pure bino. We do not apply constraints from B-physics since the colored sparticles that contribute to these processes are decoupled in our analysis.

It has been emphasized earlier in the Section 1 that our focus in this paper is to investigate the possible production and subsequent detection of electroweakinos at the LHC, relevant to the resolution of the \gmu \, anomaly. The composition of neutralinos play an important role for that purpose. Hence for the rest of the paper we have conducted separate analyses, based on the decomposition of the lightest and the second lightest neutralino, due to wide variation in the production cross-section of wino, higgsino and their admixture. To implement this we divide our parameter space into the following three regions:

\begin{align}
 M_2/\mu\geq 2, \ &&   {\rm(Region-I)} \label{higgsino} \\
 M_2/\mu\leq 0.2, \ && {\rm(Region-II)}  \label{wino} \\
 0.2<M_2/\mu<2, \ && {\rm (Region-III)}  \label{wino-higgsino}
\end{align}

 The composition of the neutralinos in each case will depend on the parameter $M_1$. In the Table~\ref{neu-comp} we therefore highlight regions of the parameter space based on the relative order of  $M_1, M_2$ and $\mu$. We discuss these cases in more detail in subsequent sections.

\begin{table}[!htp]
\begin{center}
\begin{tabular}{|c | c |c | c |} 
\hline
\hline 
 
  Region &  &  $\neu{1}$ &  $\neu{2}$ \\
\hline

\multirow{3}{*}{I} & $M_1 \gg \mu$ & higgsino & higgsino \\
                   & $M_1 \ll \mu$ &  bino & higgsino    \\
                   & $M_1 \sim \mu$ & bino-higgsino & bino-higgsino \\
\hline                   
\multirow{5}{*}{II} & \multirow{3}{*}{$M_1 \gg M_2$} & \multirow{3}{*}{wino} & bino ($M_1 \ll \mu$) \\
                   &                &       & higgsino ($M_1 \gg \mu$)\\
                   &                &       & bino-higgsino ($M_1 \sim \mu$)\\
 \cline{3-4}                    
                   & $M_1 \ll M_2$ &  bino & wino    \\
                   & $M_1 \sim M_2$ & bino-wino & bino-wino \\ 
 
\hline
\multirow{3}{*}{III} & $M_2 \sim \mu \ll M_1$ & wino-higgsino & wino-higgsino \\
                   & $M_2 \sim \mu \gg M_1$ & bino  & wino-higgsino    \\
                   & $M_2 \sim \mu \sim M_1$ & mixed  &  mixed   \\

\hline  
\end{tabular}
\end{center}
\caption{Composition of $\neu{1}$ and $\neu{2}$ in different regions of the parameter space based on the ratios $M_1/M_2$ and $M_1/\mu$.}
\label{neu-comp}
\end{table}


%

\section{Bounds on the electroweakino and smuon masses}\label{results-1}

\begin{figure}[!htp]
\centering
\includegraphics[scale=.4]{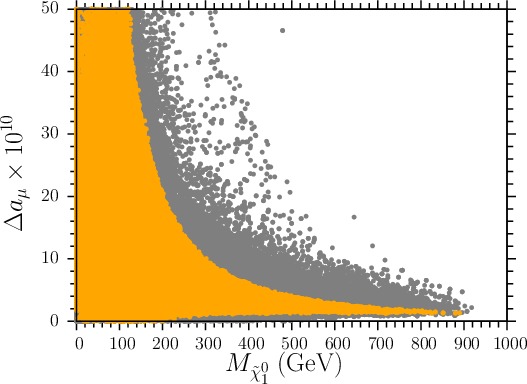}
\includegraphics[scale=.4]{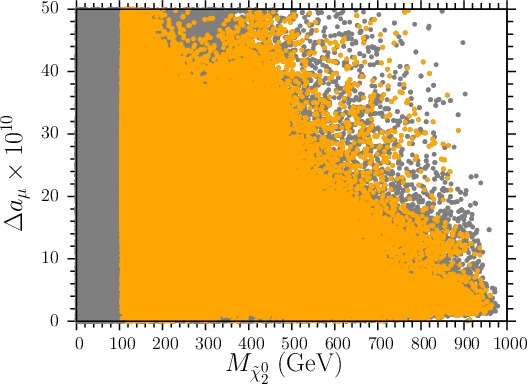}
\includegraphics[scale=.4]{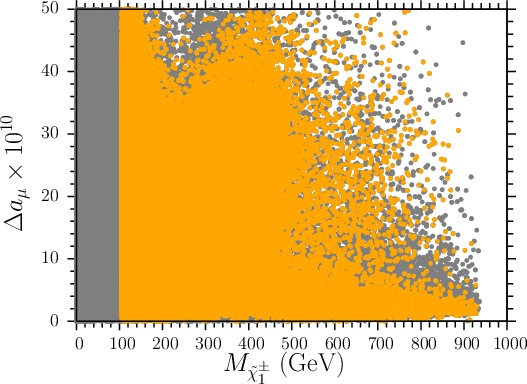}
\includegraphics[scale=.4]{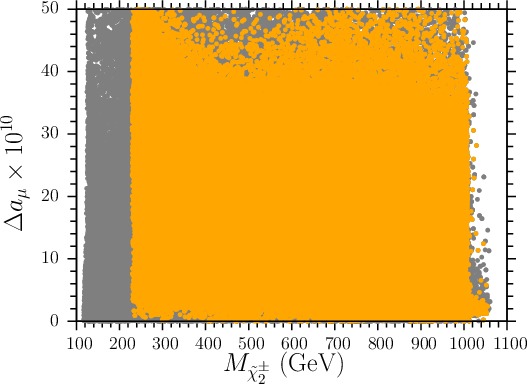}
\includegraphics[scale=.4]{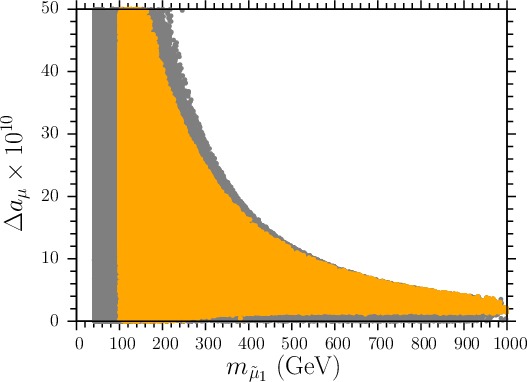}
\includegraphics[scale=.4]{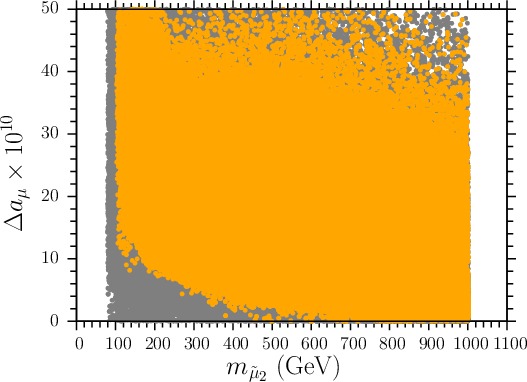}
\caption{Plots in the  $\Delta a_{\mu}-M_{\tilde{\chi}^{0}_{1}} $, $\Delta a_{\mu}-M_{\tilde{\chi}^{0}_{2}} $, $\Delta a_{\mu}-M_{\tilde{\chi}^{\pm}_1} $, $\Delta a_{\mu}-M_{\tilde{\chi}^{\pm}_2} $, $\Delta a_{\mu}-m_{\tilde{\mu}_1} $ and $\Delta a_{\mu}-m_{\tilde{\mu}_2} $ planes for  $\tan\beta=10$. \textit{Gray} points represent raw data. \textit{Orange} points are subset of the \textit{gray} points and satisfy the sparticle mass constraints given in equation (\ref{mass-constraints}).}
\label{fig-tb10-1}
\end{figure}

In this section we discuss our results for $\tan \beta =10$. For a fixed value of $\tan\beta$, the masses of the neutralinos ($M_{\tilde{\chi}^{0}_{1,2}}$), charginos ($M_{\tilde{\chi}^{\pm}_{1,2}}$) and the smuons ($ m_{\tilde{\mu}_{1}}, \, m_{\tilde{\mu}_{2}}$)\footnote{Here $ m_{\tilde{\mu}_{1,2}}$ are the mass eigenvalues of the smuon mass-matrix. From here on we have used $ m_{\tilde{\mu}_{1,2}}$ as smuon masses but returned to $ m_{\tilde{\mu}_{L,R}}$ ($ m_{\tilde{l}_{L,R}}$) notation on occasions, when distinction between left-handed and right-handed smuons (sleptons) is needed.} can affect the value of $\gmu$. We therefore show $\Delta a_{\mu}$ as a function of these parameters in Figure \ref{fig-tb10-1}. Our results are presented in the  $\Delta a_{\mu}-M_{\tilde{\chi}^{0}_{1}} $, $\Delta a_{\mu}-M_{\tilde{\chi}^{0}_{2}} $, $\Delta a_{\mu}-M_{\tilde{\chi}^{\pm}_1} $, $\Delta a_{\mu}-M_{\tilde{\chi}^{\pm}_2} $, $\Delta a_{\mu}-m_{\tilde{\mu}_1} $ and $\Delta a_{\mu}-m_{\tilde{\mu}_2} $ planes. The \textit{gray} points represent raw data and are consistent with neutralino as the LSP. \textit{Orange} points form subset of the \textit{gray} ones and satisfy the sparticle mass constraints presented in equation (\ref{mass-constraints}). As expected, we can see from Figure \ref{fig-tb10-1} that a significant region of the parameter space resolves the $\gmu$ anomaly. The $\Delta a_{\mu}-M_{\tilde{\chi}^{0}_{1}} $ plane shows a large enhancement for low values of the $\neu{1}$ mass. For the central value $\Delta a_{\mu} \simeq 28.6 \times 10^{-10}$, the upper bound on the neutralino mass is around 200 GeV.  For the lower bound on $\Delta a_{\mu}$ given in equation (\ref{gm2-2s}) the upper bound on the $\neu{1}$ mass is relaxed to $\sim$ 300 GeV.

{From the $\Delta a_{\mu}-m_{\tilde{\mu}_1} $ plane we can observe a similar large enhancement for low values of the smuon mass. For the central value of $\Delta a_{\mu}$ the upper bound on the lighter smuon mass is around 300 GeV. Again for the lower bound on $\Delta a_{\mu}$ the upper bound on the smuon mass is relaxed to $\sim$ 500 GeV. The heavier smuon mass is not bounded as can be seen from the $\Delta a_{\mu}-m_{\tilde{\mu}_2} $ plane. Note that the A terms in our analysis are set equal to zero, which implies that the physical and gauge eigenstates of the smuons are essentially the same (except for large values of $\mu$ when the mixing terms can be large). The conclusions for the left and right handed smuon masses are therefore similar to what we have concluded for the physical masses from the $\Delta a_{\mu}-m_{\tilde{\mu}_1} $ and $\Delta a_{\mu}-m_{\tilde{\mu}_2} $ planes}.

\begin{figure}[]
\centering
\includegraphics[scale=.4]{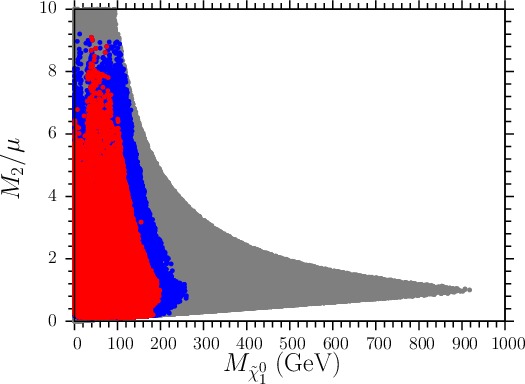}
\includegraphics[scale=.4]{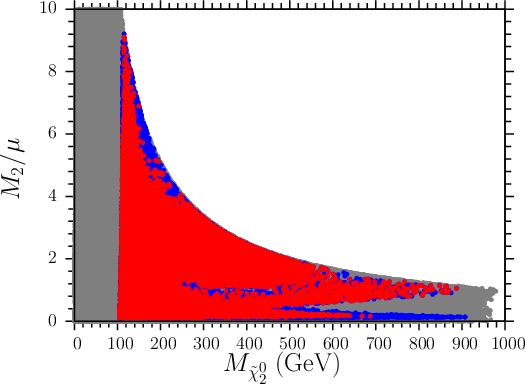}
\includegraphics[scale=.4]{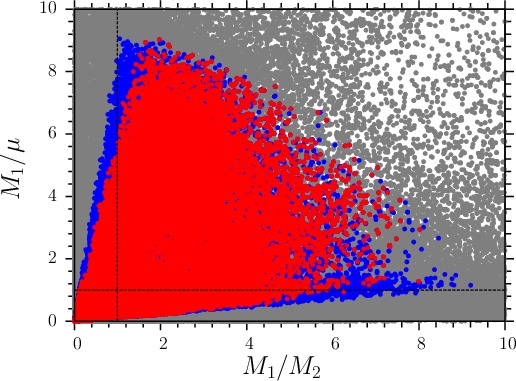}
\includegraphics[scale=.4]{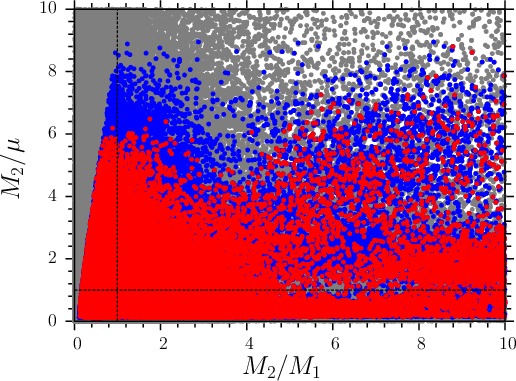}
\caption{ Plots in the $M_2/ \mu-M_{\tilde{\chi}^{0}_{1}}$, $M_2/ \mu-M_{\tilde{\chi}^{0}_{2}}$, 
$M_1/ \mu-M_1/M_2$ and $M_2/ \mu-M_2/M_1$ planes for $\tan\beta=10$. Gray points represent raw data. Blue points form subset of the gray and satisfy $\gmu$ in the $2 \sigma$ range. Red  points satisfy $\gmu$ in the $1 \sigma$ range. Red and blue points also satisfy the sparticle mass constraints given in equation (\ref{mass-constraints}).}
\label{fig-tb10-2}
\end{figure}

{As described earlier, our aim is to highlight the composition of the neutralinos that resolves the $\gmu$ anomaly.} For this purpose, Figure \ref{fig-tb10-2} displays our results in the 
$M_2/ \mu-M_{\tilde{\chi}^{0}_{1}}$, $M_2/ \mu-M_{\tilde{\chi}^{0}_{2}}$, 
$M_1/ \mu-M_1/M_2$ and $M_2/ \mu-M_2/M_1$ planes. \textit{Gray} points represent raw data and are consistent with LSP neutralino. \textit{Blue} points form  a subset of the \textit{gray} and satisfy the 2$\sigma$ deviation in $\gmu$ given in equation (\ref{gm2-2s}). Similarly, the \textit{red} points  satisfy the 1$\sigma$ deviation in $g-2$ given in equation (\ref{gm2-1s}). We can see from the $M_2/ \mu-M_{\tilde{\chi}^{0}_{1}}$ plane that insisting on 2$\sigma$ limit on $\Delta a_{\mu}$ implies that the neutralino has to be lighter than $\sim$ 260 GeV. This reduces to $\sim$ 200 GeV for the 1$\sigma$ limit. The limits on $\Delta a_{\mu}$, however, do not yield a bound on the ${\tilde{\chi}^{0}_{2}}$ mass as the $M_2/ \mu-M_{\tilde{\chi}^{0}_{2}}$ plane shows. 

The lower left and right panels of Figure \ref{fig-tb10-2} clearly show that different type of neutralino compositions are possible in this parameter space. For the region $M_1/\mu < 1 $ and $M_1/M_2< 1 $, the LSP is expected to be essentially a pure bino. As is well known, a pure bino type $\neu{1}$ yields a large relic abundance since the cross sections involved are small. However, coannihilation of the bino with other sparticles can resolve this issue. Moreover, the correct relic abundance can also be achieved if the $\neu{1}$ acquires a wino or a higgsino component. This is possible in our analysis since the region $M_1/\mu < 1$ and $M_1/M_2 > 1 $ corresponds to a mixed bino-wino type $\neu{1}$. Similarly, the region $M_1/\mu > 1$ and $M_1/M_2 < 1 $ corresponds to a bino-higgsino type $\neu{1}$.

 From the lower right panel of Figure \ref{fig-tb10-2} we can see from the unit lines that the $\neu{1}$ can essentially be a pure wino for a notable region of the parameter space corresponding to $M_2/\mu < 1 $ and $M_2/M_1< 1 $. The lower panels of Figure \ref{fig-tb10-2} further show that the $\neu{1}$ can also be a pure higgsino for  $M_1/\mu > 1 $ and $M_2/\mu > 1  $. These plots therefore show that a pure bino, wino and higgsino can satisfy the 2$\sigma$ limit on $\Delta a_{\mu}$. Plots for different regions, as defined in equations (\ref{higgsino}), (\ref{wino}), (\ref{wino-higgsino}), are shown separately when they are discussed in detail in Section~\ref{results-2}.

\begin{figure}[]
\centering
\includegraphics[scale=.4]{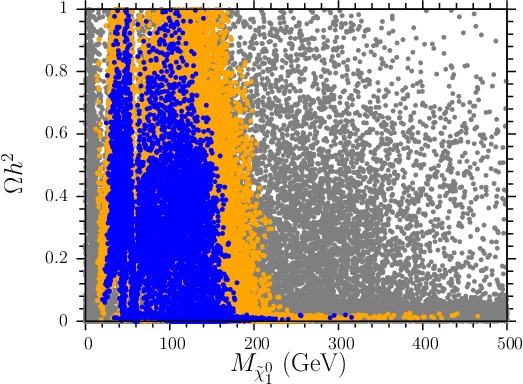}
\includegraphics[scale=.4]{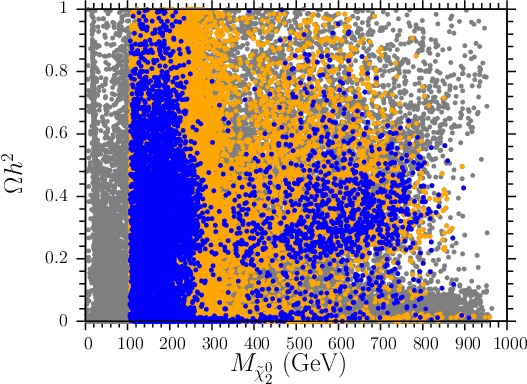}
\includegraphics[scale=.4]{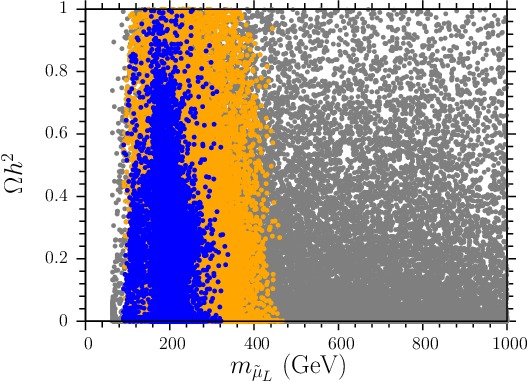}
\caption{ Plots in the $ \Omega h^2-M_{\tilde{\chi}^{0}_{1}}$, $ \Omega h^2-M_{\tilde{\chi}^{0}_{2}}$ and
$ \Omega h^2-m_{\tilde{\mu}_{L}}$ planes for the $\tan\beta=10$ case. Gray point are raw data. Orange points satisfy $\gmu$ within $2 \sigma$ and blue satisfy $\gmu$ within the $1 \sigma$ range. The relic density was calculated using micromegas. Orange and blue points also satisfy the sparticle mass constraints given in equation (\ref{mass-constraints}).}
\label{fig-rd}
\end{figure}

  It is well known that $\neu{1}$ LSP is a promising candidate for weakly interacting massive particle (WIMP) DM. In Figure \ref{fig-rd} we display the relic density plots in the
$ \Omega h^2-M_{\tilde{\chi}^{0}_{1}}$, $ \Omega h^2-M_{\tilde{\chi}^{0}_{2}}$ and
$ \Omega h^2-m_{\tilde{\mu}_{L}}$ planes. The relic density was calculated using \micromegas~\cite{Belanger:2001fz}. As before, the \textit{orange} and \textit{blue} points satisfy the sparticle mass constraints given in equation (\ref{mass-constraints}). We can see that the relic density bound can be easily satisfied in this case owing to the mixed nature of the lightest neutralino and also due to  neutralino-smuon coannihilation in this scenario. However we are not confined to the relic density allowed regions for our $\gmu$ analysis.

 We find that Figures~\ref{fig-tb10-1}-\ref{fig-rd} do not change significantly for the $\tan \beta = 50$ case. Benchmark points (BP) for $\tan \beta = 10$ and 50 are shown in Tables \ref{tab1} and \ref{tab2}. In these tables we display the maximum values of the masses (in GeV) of smuons, neutralinos and charginos for $\tan\beta=10$ and $50$. The values presented in each column correspond to $\gmu$ within 1$\sigma$ and those in the brackets are for $\gmu$ within 2$\sigma$.  We should point out here that for both Region-I and Region-II, $M_{\neu{2}} > 1$ TeV can still satisfy $\gmu$ at the $2\sigma$ level, but they are not shown in aforementioned figures and tables since we have scanned the parameter space for $M_2$ and $\mu$ only upto 1 TeV. However, we have discussed these scenarios ($M_{\neu{2}} > 1$ TeV) in subsequent Sections, whenever they are relevant for the collider study.

\newpage

\begin{table}[t!]\vspace{1.5cm}
\centering
\begin{tabular}{|p{2cm}|p{3.5cm}p{3.5cm}p{3.5cm}|}
\hline
\hline
                 	&	 Region-I 	&	 Region-II 	&	 Region-III 	\\
						
\hline

$m_{\tilde{\mu}_1}$         	&$	298.63	\	(	426.84	)	$&$	227.97	\	(	306.81	)	$&$	338.01	\	(	470.12	)	$\\
$m_{\tilde{\mu}_2}$         	&$	1000.67	\	(	1000.79	)	$&$	999.40	\	(	999.61	)	$&$	1000.58	\	(	1000.73	)	$\\
$m_{\tilde{\chi}^0_1}$         	&$	163.94	\	(	218.31	)	$&$	178.63	\	(	192.12	)	$&$	198.03	\	(	259.97	)	$\\
$m_{\tilde{\chi}^0_2}$         	&$	488.30	\	(	488.30	)	$&$	687.26	\	(	907.32	)	$&$	886.93	\	(	886.93	)	$\\
$m_{\tilde{\chi}^\pm_1}$      	&$	487.54	\	(	487.54	)	$&$	196.22	\	(	196.22	)	$&$	886.94	\	(	886.94	)	$\\
$m_{\tilde{\chi}^\pm_2}$       	&$	1008.63	\	(	1008.63	)	$&$	1006.75	\	(	1006.75	)	$&$	1029.14	\	(	1029.14	)	$\\

\hline		  		  		  	

\hline
\end{tabular}
\caption{ Maximum values of the masses of smuons, neutralinos and charginos for $\tan\beta=10$, resulted from our MSSM parameter scan. The values presented in each column correspond to $\gmu$ within $1\sigma$ and those in the brackets are for $\gmu$ within $2 \sigma$. All the masses are in GeV.}
\label{tab1}
\end{table}


\begin{table}[t!]\vspace{1.5cm}
\centering
\begin{tabular}{|p{2cm}|p{3.5cm}p{3.5cm}p{3.5cm}|}
\hline
\hline
                 	&	 Region-I 	&	 Region-II 	&	 Region-III 	\\
			
\hline

$m_{\tilde{\mu}_1}$         	&$	911.78	\	(	992.38	)	$&$	715.16	\	(	904.52	)	$&$	957.74	\	(	996.20	)	$\\
$m_{\tilde{\mu}_2}$         	&$	1000.88	\	(	1000.88	)	$&$	1000.92	\	(	1000.92	)	$&$	1000.86	\	(	1000.93	)	$\\
$m_{\tilde{\chi}^0_1}$         	&$	390.05	\	(	478.42	)	$&$	197.71	\	(	197.71	)	$&$	482.08	\	(	637.11	)	$\\
$m_{\tilde{\chi}^0_2}$         	&$	477.76	\	(	491.96	)	$&$	963.56	\	(	963.56	)	$&$	947.02	\	(	966.17	)	$\\
$m_{\tilde{\chi}^\pm_1}$      	&$	477.19	\	(	487.97	)	$&$	197.95	\	(	197.95	)	$&$	910.37	\	(	939.92	)	$\\
$m_{\tilde{\chi}^\pm_2}$       	&$	1007.69	\	(	1007.89	)	$&$	1006.65	\	(	1006.65	)	$&$	1033.37	\	(	1055.87	)	$\\

\hline		  		  		  	

\hline
\end{tabular}
\caption{ Maximum values of the masses of smuons, neutralinos and charginos for $\tan\beta=50$, resulted from our MSSM parameter scan. The values presented in each column correspond to $\gmu$ within $1\sigma$ and those in the brackets correspond to $\gmu$ within $2\sigma$. All the masses are in GeV.}
\label{tab2}
\end{table}

\begin{figure}[!t]
\centering

\includegraphics[scale=.3]{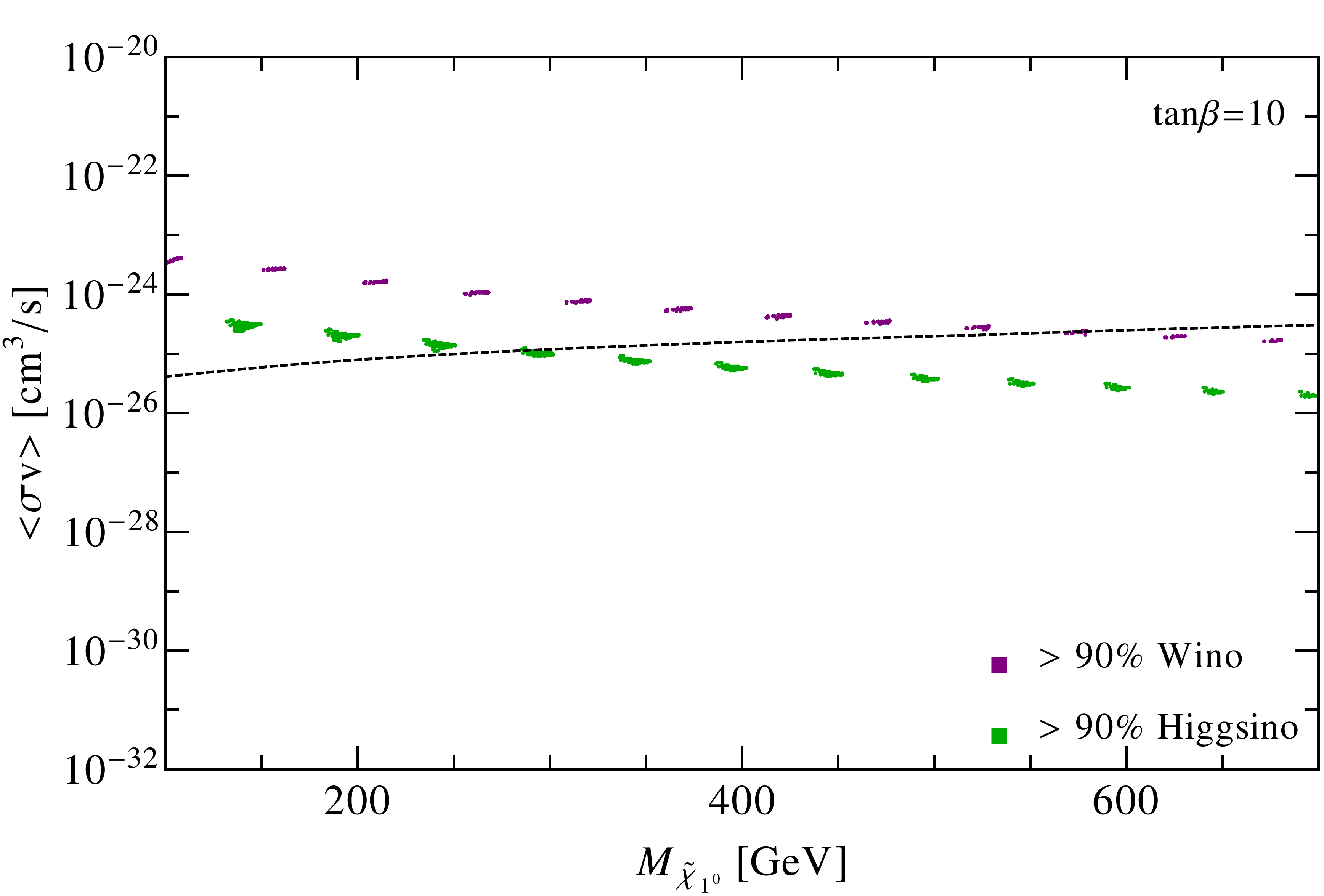}
\caption{Constraints on the annihilation cross-section into $WW$+($ZZ$) final states and mostly wino/higgsino annihilation cross-section as a function of neutralino mass. The black dot-dashed curve is the constraint from the photon spectrum of dSPhs assuming NFW DM profile. The Fermi-LAT bounds in the $WW$+($ZZ$) channel is extracted by digitizing the Figure 8 of Ref.~\cite{Fermi:pass8}.}
\label{fig-indirect}
\end{figure}

  $\neu{1}$ can self-annihilate into the Standard Model (SM) particles (quarks, leptons, $W,Z,h$-bosons etc). WIMPs are being searched indirectly, by different astrophysical experiments, through the particles (proton, neutrinos, photon) these quarks, leptons and $W,Z,h$-bosons produce in turn. The Fermi-LAT collaboration is one such experiment, which provides stringent bounds on DM annihilation cross-section from their study of the gamma-ray spectrum from dwarf spheroidal galaxies (dSphs) of the Milky Way~\cite{Ackermann:2013yva,Fermi:pass8}. Ref.~\cite{Fan:2013faa} has studied these constraints arising from dSphs, in the context of neutralino DM and ruled out wino DM upto 385 GeV and higgsino DM upto 160 GeV using $WW$+($ZZ$) annihilation channel. We have scanned our parameter space for the same, with newly released Pass8 data by Fermi-LAT~\cite{Fermi:pass8} and upgrade the results of Ref.~\cite{Fan:2013faa}. The Fermi-LAT bounds in the $WW$+($ZZ$) channel is extracted by digitizing the Figure 8 of Ref.~\cite{Fermi:pass8}. We found that assuming NFW DM profile, mostly wino type ($\geq 90 \% $) $\neu{1}$ is ruled out upto $\sim 575$ GeV, while mostly higgsino type LSP is ruled out upto $\sim 275$ GeV. Mostly bino-type $\neu{1}$ remain unconstrained from dSphs.The results of our scan is presented in Figure~\ref{fig-indirect} for $\tan \beta = 10$. The conclusion remain the same for $\tan \beta=50$. The annihilation cross-sections are calculated by using  \micromegas~\cite{Belanger:2001fz}. 

  However it is well established that for a pure wino or higgsino-type $\neu{1}$, the observed DM relic density can not be satisfied for
 $M_{\neu{1}}$ less than $\sim 2.5$ TeV for wino and $\sim 1$ TeV for higgsinos, due to their large annihilation cross-sections~\cite{well-temp}. Hence for the mass range of the LSP allowed by $\gmu$, we require additional component of DM (axion is a possible candidate~\cite{axion-Baer,axion-Baer2}) to saturate the relic density. If the composition of the DM remains the same since the thermal freeze-out, the constraint on the annihilation cross-section of the LSP, coming from dSphs, will be relaxed substantially due to reduced WIMP abundance.  In addition if we remove the ultra-faint dwarf galaxies and restrict ourselves to eight classical dwarfs only then the indirect detection limits weaken by a factor of $\sim 2$ for $m_{\chi} \gtrsim 500$ GeV, but the impact on the combined limits for soft annihilation spectra with $m_{\chi} \lesssim 100$ GeV is only $\pm 20 \%$~\cite{Ackermann:2013yva}.
 
    The DM direct detection searches can also impose strong constraint on the LSP mass, especially on bino-higgsino like $\neu{1}$~\cite{DMDD-BlindSpot,g-2vDD,higgsino-DD}. Ref.~\cite{g-2vDD} points towards a tension between $\gmu$ allowed parameter space and XENON100 results, but, under GUT inspired universality condition, $M_1=0.5M_2$. However these bounds require precise knowledge about the properties of proton and may vary by a factor of 3 due to uncertainties involved in the available data~\cite{Arnowitt}. Moreover these bounds can be occasionally evaded with correct assignment of sign for the gaugino and higgsino mass parameters. In that case the direct detection cross-sections get suppressed due to fortuitous cancellations between contributions from different SUSY Higgs eigenstates, as shown in Ref.~\cite{DMDD-BlindSpot,Arnowitt,neg-mu-DD}. We should recall here that the contribution to $\gmu$ is largest when $M_1,M_2$ and $\mu$ possess the same sign and hence assigning opposite signs to gaugino and higgsino parameters is not favoured by $\gmu$. In contrast, setting $m_A$ to be light~\footnote{ We have set $m_A$ to be heavy since they don't contribute towards $\gmu$ at 1-loop level. However the CP-odd Higgs, $A$, do contribute to $\gmu$ at 2-loop level by means of Barr-Zee diagrams~\cite{Barr-Zee} but their contribution is small in the parameter space we considered~\cite{Barr-Zee-MSSM}.} may give rise to additional blind-spots in direct detection limits~\cite{Huang:mA} but then also a sizeable part of parameter space we studied for collider will be ruled out by Br($B_s \rightarrow \mu^+ \mu^− $) and Br($b \rightarrow s\gamma$) constraints. Finally, the direct detection bounds for mostly higgsino-type DM are redundant if we consider depleted DM abundance of higgsinos~\cite{axion-Baer2}.
  
  Thereby we did not impose any DM constraints on the parameter space we scanned for this study. However, if the constraints are applied, the LHC reach can easily be obtained from the tables we shall provide in the next two sections.


%

\section{\label{LHC} Production of electroweakinos at the LHC}

In this section we shall discuss the production  of electroweakinos at the LHC, pertaining to the parameter space considered in the previous sections. The LHC experiments (CMS and ATLAS) have set fairly stringent lower limits [$\mathcal{O}$(TeV)] on the squarks ($\tilde{q}$) and gluino ($\tilde{g}$) masses~\cite{:2012rz, Aad:2012hm,cmssusy, :2012mfa}. Hence the production of electroweakinos via cascade decays of $\tilde{q}$ and $\tilde{g}$ has been neglected, and we focus on the pair production of electroweakinos by Drell-Yan (DY) processes, in association with radiated jets:  
\begin{align}
p p \rightarrow \neu{k} \neu{l} j, \neu{k}  \chpm{l} j,  \chpm{k}  \chmp{l}  j,
\end{align}
where $k,l=1,2,3,4$ for neutralinos,  $k,l=1,2$ for charginos, and $j$ denotes the hadronic jets. Wino-like and higgsino-like electroweakinos will be sufficiently produced by this mechanism at the LHC, owing to their large couplings to $W,Z$ and $\gamma$. Due to unsuppressed $SU(2)_{L}$ couplings, electroweakino pair production by $W$-exchange will have the largest production cross-section, while the contribution from $t$-channel squark exchange diagrams is negligible due to heavy squark masses.

The electroweakinos can also be produced by Vector Boson Fusion (VBF) processes but the production cross-section is small in those channels. However, VBF, characterised by two highly energetic forward jets in opposite hemisphere and large $\met$, can be complementary to DY processes in probing the EW structure of MSSM. VBF processes can also be very useful in probing small mass-gap scenarios due to their highly boosted topology, as shown in Refs.~\cite{VBF-ewk,VBF-sl}. 

 The signal samples are generated upto $\mathcal{O}(\alpha^{4}_{EW} \alpha^{4}_s)$ and include 1-parton (inclusive) processes. $(t \rightarrow b  l \nu) \,\bar{t} \, + \,$ jets, $ (W \rightarrow l \nu) \, W \, + \,$ jets, $ (W \rightarrow l \nu) \, Z \, + \,$ jets, $ZZ \, + \,$ jets, $(W \rightarrow l \nu) \, + \,$ jets and $ (Z \rightarrow l l) \, + \,$ jets, where $l = e, \mu, \tau$, are the SM backgrounds considered for all the studies presented in this paper. The $VV \, + \,$ jets (where $V = W, \, Z$) background consists of up to 2-partons inclusive processes, while the $\ttbar \, + \,$ jets and $V \, + \,$ jets include up to 3-partons inclusive processes. The MLM-scheme for jet matching~\cite{Mangano:2006rw} is used to avoid double-counting. 

 The signal and background samples, used in this paper, are generated with \madgraph \, \cite{Alwall:2011uj}. These samples are then passsed through \pythia \, \cite{Sjostrand:2006za} for parton showering and hadronization, and finally through \pgs \, \cite{pgs} to simulate the effect of detectors. The $\ttbar+ \,$ jets and $VV \, + \,$ jets, which are dominant backgrounds for multi-lepton $ + \, \met$ final states, are scaled to  NLO values by using the K-factor presented in Ref.~\cite{ttbar-NLO} and Ref.~\cite{VV-NLO} respectively.

\section{\label{results-2} Results}
 
 We have chosen several BPs for analysis from the parameter spaces discussed in section~\ref{g-2}. As previously mentioned, the SUSY parameters are selected such that the coloured super partners are sufficiently heavy along with all Higgs particles except for the lightest (SM-like) Higgs.  We also set the masses of the left-handed and right-handed sleptons to be the same in order to maximize the BR for $\neu{2} \rightarrow \tilde{l} l$ decay in the $m_{\tilde{l}} < M_{\neu{2}}$ case. Next we discuss the results for each of the regions described in Eqs.~\ref{higgsino},\ref{wino},\ref{wino-higgsino}. Each of these regions are divided into sub-regions depending on the nature of the LSP. For simplicity we restricted ourselves to $M_1 < min(M_2,\mu)$ and $M_1 > max(M_2,\mu)$ only. Hence each region contain two sub-regions corresponding to bino-like LSP and non bino-like (wino, higgsino or wino-higgsino) LSP. For $M_1 > max(M_2,\mu)$ cases we set $M_1=1$ TeV. Next the BPs, with $\tan \beta=$ 10 and 50, are classified into the following two broad classes due to different search strategies needed at the LHC to probe them:
\begin{align}
(i) \,  m_{\tilde{l}} > M_{\neu{2}} \label{slltneu2},\\
(ii) \,  m_{\tilde{l}} < M_{\neu{2}} \label{slgtneu2},
\end{align}

where $\tilde{l}=\tilde{e},\tilde{\mu}$. These two cases are further subdivided into different scenarios depending on the different mass-splittings between the neutralinos and sleptons. 

\subsection{\label{I} Region - I \, ($M_2/\mu\geq 2$)}

 In this case, with $M_2 \geq 2 \mu$, the nature of $\neu{1}$ and $\neu{2}$ is determined by the relative magnitude of $M_1$ and $\mu$. If $\mu/ M_1 \ll 1$, both $\neu{1}$ and $\neu{2}$ will be higgsino-type, and for $\mu/ M_1 \gg 1$, the LSP will be bino-type and $\neu{2}$ will be higgsino-type. Otherwise, they will be mixed states with appropriate composition. 

\begin{figure}[!ht]
\centering
\includegraphics[scale=.4]{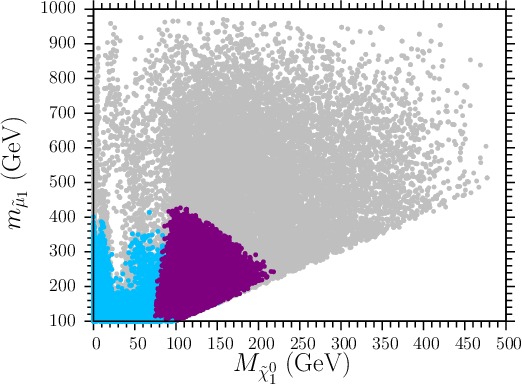}
\includegraphics[scale=.4]{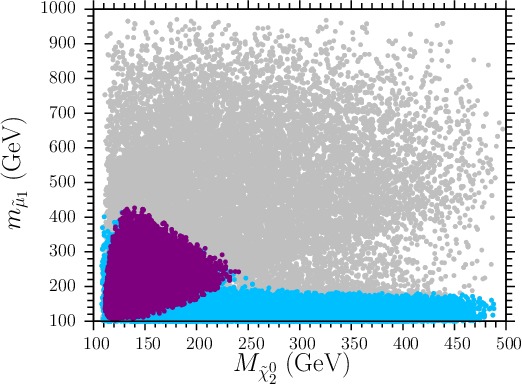}
\includegraphics[scale=.4]{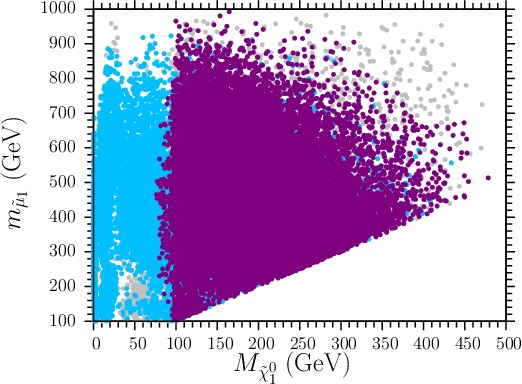}
\includegraphics[scale=.4]{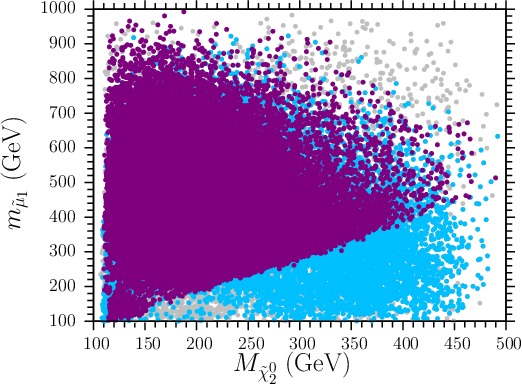}
\caption{Plots in the $m_{\tilde{\mu}_1}-M_{\tilde{\chi}^{0}_{1}}$ and $m_{\tilde{\mu}_1}-M_{\tilde{\chi}^{0}_{2}}$ planes for $\tan\beta=10$ (\textit{upper panel}) and $\tan \beta=50$ (\textit{lower panel}). All points in these plots satisfy the definition given in equation (\ref{higgsino}) for Region-I. Light gray points in this plot satisfy the constraints given in equation (\ref{mass-constraints}). Light blue points are subset of the gray, satisfy $\gmu$ in the $2 \sigma$ range and $M_1/\mu<1$. Purple points are subset of the gray, satisfy $\gmu$ in the $2 \sigma$ range and also $M_1/\mu>1$.}

\label{fig:higgsino1}
\end{figure}

In Figure \ref{fig:higgsino1} we display our results in the 
$m_{\tilde{\mu}_1}-M_{\tilde{\chi}^{0}_{1}}$ and $m_{\tilde{\mu}_1}-M_{\tilde{\chi}^{0}_{2}}$ 
planes for this region for $\tan \beta$ values of 10  (\textit{upper panel}) and 50 (\textit{lower panel}). \textit{Light gray} points satisfy the LSP neutralino constraint and also the constraints given in equation (\ref{mass-constraints}). \textit{Light blue} points are subset of the gray, and they satisfy $\gmu$ in the $2 \sigma$ range and $M_1/\mu<1$. \textit{Purple} points are subset of the gray, satisfy $\gmu$ in the $2 \sigma$ range and also $M_1/\mu>1$. For this case $\tilde{\mu}_1$ has to be lighter than $\sim$ 400 GeV with $\tan \beta = 10$. For the \textit{purple} points, the $\neu{1}$ will essentially be a pure higgsino, whereas for the \textit{light blue} points the $\neu{1}$ will acquire a bino component. The pure higgsino $\neu{1}$ mass lies in the range:
70 GeV $\lesssim M_{\tilde{\chi}^{0}_{1}} \lesssim $ 200 GeV. Note also that this bound is narrower for relatively heavier $\tilde{\mu}_1$, i.e., for $m_{\tilde{\mu}_1} \simeq$ 400 GeV, $M_{\tilde{\chi}^{0}_{1}} \simeq 100$ GeV. The mixed bino-higgsino $\neu{1}$ (\textit{blue} points) can be much lighter in this case. The mass of the heavier $\tilde{\chi}^{0}_{2}$ is also bounded in this case, namely 100 GeV $\lesssim M_{\tilde{\chi}^{0}_{2}} \lesssim $ 250 GeV. 
It should be noted that the above results are for $\tan \beta = 10$, while for $\tan \beta = 50$ a wider range of smuon and neutralino masses satisfy \gmu.  

 We subdivide this section into sub-sections depending on the nature of the LSP. For simplicity we have restricted ourselves only to pure bino and pure higgsino like scenarios. However, as previously mentioned in  Section~\ref{results-1}, bino-higgsino as a LSP candidate is strongly disfavoured by direct detection experiments. We set $M_2/\mu = 2$ for subsequent collider studies.
  
\subsubsection{Bino LSP}

Due to very small production rate of bino at the LHC it is futile to search for them directly~\cite{VBF-DM}. Hence we concentrate on searching for heavier neutralinos and charginos for bino-like LSP scenarios.

\subsubsection*{$\mathbf{Case \, (i): m_{\tilde{l}} > M_{\neu{2}}}$\footnote{This also ensures $m_{\tilde{l}} > M_{\neu{3}}$ for higgsino-type $\neu{2,3}$.}}

\begin{table}[t]
\begin{center}
\begin{tabular}{|c | c |c | c | c | c |} 
\hline
\hline 
  Region & $M_2/\mu$ & $M_{\neu{1}}$ & $\tan \beta$ & $M_{\neu{2}}$ &  Significance($\sigma$) \\ 

         &    & [GeV] &   & [GeV]  &  $\dfrac{S}{\sqrt{(S+B)}}$ \\
\hline
\multirow{4}{*}{I} & \multirow{4}{*}{2} & \multirow{14}{*}{150} & 10 & 200  & 30.0 \\
 \cline{4-6}
  &  &  & \multirow{3}{*}{50} & 300 & 18.6 \\
  &  &  &                     & 400 & 12.3  \\
  &  &  &                     & 500 & 8.38   \\
\cline{1-2}\cline{4-6}

\multirow{5}{*}{II} & \multirow{5}{*}{0.2} &  & \multirow{2}{*}{10} & 200  & 123 \\
  &  &  &  & 250 &  6.55 \\
 \cline{4-6}
  &  &  & \multirow{3}{*}{50} & 300 & 3.14 \\
  &  &  &                     & 400 & 6.17  \\
  &  &  &                     & 500 & 4.55   \\ 
\cline{1-2}\cline{4-6}

\multirow{5}{*}{III} & \multirow{5}{*}{0.75} &  & \multirow{2}{*}{10} & 200  & 52.6 \\
  &  &  &  & 250 &  19.2 \\
  \cline{4-6}
  &  &  & \multirow{3}{*}{50} & 350 & 12.8 \\
  &  &  &                     & 500 & 7.74  \\
  &  &  &                     & 600 & 5.32  \\
  \hline

\end{tabular}
\end{center}
\caption{Significances at the LHC at $\sqrt{s}=14$ TeV and 3000 fb$^{-1}$ of integrated luminosity for $m_{\tilde{l}} > M_{\neu{2}}$. Benchmark points with different $M_2/ \mu$ values, belonging to different regions as defined in equations (\ref{higgsino}), (\ref{wino}), (\ref{wino-higgsino}) with Bino-type LSP, are presented which satisfy \gmu \, requirement and are also not excluded by 8 TeV LHC results. See text for details. }
\label{slgtneu2Reach}
\end{table}

 The LHC experiments are pursuing the search for electroweakinos and sleptons in various final states, and the non-observation of any signal in Run-I has already provided impressive lower bounds on the masses of these particles. The conventional multi-lepton plus $\met$ channels are followed by ATLAS and CMS~\cite{ATLASS2lep,ATLASS3lep,ATLASewkinoWh,CMSSlep1,CMSSewkinoWh}, and the current bound on $M_{\neu{2}}$ is $\sim 425$ GeV (for $M_{\neu{1}}=0$ GeV) in $WZ$ channel~\cite{ATLASS2lep}. However, these bounds are derived under highly simplified assumption that $\neu{2},\chpm{1}$ decays into gauge bosons with a 100$\%$ branching ratio (BR), while in scenarios pertaining to $\gmu$, BR($\neu{2} \rightarrow \neu{1} Z$) is $\lesssim 30 \%$ for $M_{\neu{2}} - M_{\neu{1}} > m_h$. These scenarios are dominated by $\neu{2} \rightarrow \neu{1} h$ decay. More importantly these bounds are derived by the LHC experiments assuming bino-type $\neu{1}$ and wino-type $\neu{2}$. The wino production cross-section is $\sim 3-4$ times larger than that of higgsino but higgsino signal can be augmented by the presence of light $\neu{3}$. Consequently these bounds are comparatively weaker than those limits quoted above.
 
 For $Wh$ final state the bounds are much weaker. ATLAS~\cite{ATLASewkinoWh} offers the strongest bound of $M_{\neu{2}} \sim 270$ GeV (with $M_{\neu{1}}=0$ GeV). Nonetheless the bounds are non-existent for $M_{\neu{1}} > 150$ GeV, except for $300 < M_{\neu{2}} < 400$ GeV in $WZ$ channel~\cite{ATLASS2lep}, again with the assumption of 100$\%$ BR. Taking all these bounds into account we have set $M_{\neu{1}}=150$ GeV for our BPs. However, in order to ensure that our BPs are not excluded, we have confirmed their viability with the observed results of Refs.~\cite{ATLASS2lep,ATLASS3lep,CMSSlep1} using the package \CheckMATE~\cite{CheckMATE}.
 
 This case is further classified into two sub-cases based on different values of the mass-gap $\Delta m = M_{\neu{2}}-M_{\neu{1}}$. Since the sleptons are heavier than $\neu{2}$, this region is characterised by $\neu{2} \rightarrow \neu{1} \, Z/h$ decay, and it can be subdivided depending on whether the $Z/h$ bosons produced are on-shell or off-shell. Consequently, we have chosen two class of benchmark points, namely for $\Delta m=50$ GeV and $\Delta m \geq m_Z$  respectively. 

{\bf $\Delta m=50$ GeV -} 
 Probing the small mass gap scenarios has proved to be challenging for the LHC experiments due to the difficulty in detecting the soft leptons~\cite{ATLASS2lep,ATLASS3lep,CMSSlep1}. The region $M_{\neu{2}} - M_{\neu{1}} < m_Z$ remains unconstrained by these experiments. Although various search strategies for probing $\Delta m \sim 1-50$ GeV have been proposed in the literature ~\cite{CompressedGaugino,Han,Baer:2014kya,Han:2013usa,Gori:2013ala}, we have restricted ourselves to $\Delta m \gtrsim 50 $ GeV. Our search strategy for this scenario in $3 \, l + 1 \, j + \met $ channel is similar to that discussed in Ref.~\cite{Gori:2013ala}. However, Ref.~\cite{Gori:2013ala} has taken into account the $WZ$ background only, we find that $t \bar{t}$ +jets is the dominant background for these scenarios and, therefore cannot be neglected. A combination of $\met$ cut ($> 50$ GeV), an upper-cut on the $p_T$ ($< 50$ GeV) of the leading lepton and selecting events with opposite sign same flavor (OSSF) di-lepton invariant mass ($M_{l^{\pm} l^{\mp}}$)~\footnote{The lower cut on $M_{l^{\pm} l^{\mp}}$ is imposed for di-lepton trigger purposes (See Ref.~\cite{Gori:2013ala} for details).}  between 12 and 50 GeV are principal kinematic discriminants for this study. The details of the search strategy for  moderately compressed scenarios are presented in Appendix~\ref{SS1}, along with cut-flow table containing the signal and background yields. The statistical significance at 14 TeV and 3000 fb$^{-1}$ integrated luminosity is $30 \sigma$ for $M_{\neu{2}} = 200$ GeV, $M_{\neu{1}} = 150$ GeV and $\tan \beta = 10$ ($\tan \beta =50$ is not allowed by \gmu \,). Note that $\sigma = S/\sqrt{S+B}$, where $S$ and $B$ are the signal and background rates respectively, has been used as a measure of statistical significance in this paper.

{\bf $\Delta m \geq m_Z$ -}   This scenario is more straightforward and we roughly follow the guidelines set by the CMS experiment~\cite{CMSSlep1} in $ 2 \,l+ \geq 2j + \met$, $3 \, l + \met$ and $4 \, l + \met$ final states with cuts being optimized for $\sqrt{s} = 14$ TeV together with additional cuts proposed in this work. The details of the search strategy are given in Appendices~\ref{SS2}, \ref{SS3} and \ref{SS5}. For $\tan \beta = 10$, $\Delta m \geq m_Z$ is not allowed by \gmu (at $2 \sigma$) for higgsino-like $\neu{2}$, but with $\tan \beta = 50$ points are allowed for $M_{\neu{2}} \sim 300-550$ GeV with $M_{\neu{1}}$ fixed at 150 GeV.  

 The possibility of testing these points at the LHC at 14 TeV are encouraging. The combined significances are $> 5 \sigma$ for these points. The $95\%$ CL will be $M_{\neu{2}} \sim 975$ GeV. Large $\met$ cuts ($> 200$ GeV) in all multi-lepton + $\met$ channels are found to be very effective in reducing the SM backgrounds. Additionally the application of $\Delta \phi (\met, l_3) > 1$, where $l_3$ is the 3rd lepton coming from $\chpm{1} \rightarrow \neu{1} W^{\pm}$ decay, and asymmetric $M_{T_2} > 250$ GeV cuts leaves the $3 \, l \, + \, \met$ channel devoid of any $\ttbar$ and $WZ$ backgrounds. The asymmetric stransverse mass, $M_{T_2}$, is computed out of the $\met$, the reconstructed $Z$-boson (OSSF lepton pair having invariant mass within 20 GeV window of $m_Z$) as the visible particle on one chain and $l_3$ on the other~\cite{Padhi}. $M_{T_2}$ algorithm of Ref.~\cite{MT2} has been adapted for the above computation. On the other hand rejection of events having transverse mass, $M_T = \sqrt{2 \met p_{T_l} (1-\cos(\Delta\phi_{l,\met}))}$, between 40 and 150 GeV reduces $\ttbar$ and $WW$ backgrounds by an order of magnitude in the $2 \, l \, + \, \met$ channel. The cut-flow table detailing the signal and background efficiencies of various cuts imposed, are also tabulated in Appendices~\ref{SS2}, \ref{SS3} and \ref{SS5}.

All the multi-lepton final states discussed above arises from $WW,\, WZ$ or $ZZ$ decay channels of the charginos and neutralinos. However $M_{\neu{2}} - M_{\neu{1}} > m_h$ points are dictated by $\neu{2} \rightarrow \neu{1} h$ decay with large BR. Although we did not consider the b-quark final states arising from $h \rightarrow b \bar{b}$ decay, but for the sake of completeness, we discuss the same-sign $2 \, l \, + \, 2/3 \,j + \met$ final state coming from $Wh \rightarrow W W W^*$ channel in the Appendix~\ref{SS6}. The potential of this channel to search for electroweakinos is limited. The most promising final state is found to be the $3 \, l \, + \, \met$ but this channel is not effective when $\Delta m$ is not significantly larger than $m_Z$. In those case the $4 \, l \, + \, \met$ channel is the dominant one due to large $\neu{2}\neu{3}$ cross-section of the higgsinos. In comparison the $2 \, l \, +  \geq2 \, j \, + \, \met$ channel suffers from low $S/B$ ratio.


The expected combined statistical significances at 3000 fb$^{-1}$ of integrated luminosity are tabulated in Table~\ref{slgtneu2Reach} (the significances of different channels are added in quadrature to obtain the combined significance.) For a complementary study we refer the reader to Ref.~\cite{Padhi}, where they have also searched for elctroweakinos with $M_{\neu{1}}=0$ GeV and included the $Wh$ and $Zh$ channels as well, in addition to multilepton channels. They report an exclusion limit of 480 GeV at $95\%$ CL, for pure higgsino-like $\neu{2}$ with 300 fb$^{-1}$ of integrated luminosity. However for our BPs with $M_{\neu{1}} \geq 150$ GeV the system has much smaller $\met$ compared to $M_{\neu{1}}=0$ case. Consequently we lack the handle that is required to suppress the $t\bar{t}$ background and we don't expect any significant improvement in significance by adding these channels as shown by Ref.~\cite{Yu}.

\subsubsection*{$\mathbf{Case \, (ii): m_{\tilde{l}} < M_{\neu{2}}}$} 

\begin{table}[!htp]
\begin{center}
\begin{tabular}{|c | c |c | c | c | c |} 
\hline
\hline 
  Region & $M_2/\mu$ & $M_{\neu{1}}$ & $\tan \beta$ & $M_{\neu{2}}$ &  Significance($\sigma$) \\ 

         &    & [GeV] &   & [GeV]  &  $\dfrac{S}{\sqrt{(S+B)}}$ \\
\hline
\multirow{9}{*}{I} & \multirow{9}{*}{2} & \multirow{4}{*}{150} & \multirow{3}{*}{10} & 400  & 52.9 \\
  &  &  &  & 500 & 31.1 \\
  &  &  &  & 700 & 7.07 \\
  \cline{4-6}
  &  &  & 50 & - & - \\
  \cline{3-6}
  &  & \multirow{5}{*}{250} & 10 & - & - \\
  \cline{4-6}
  &  &  &  \multirow{4}{*}{50} & 300 & 66.4 \\
  &  &  &  & 400 & 28.6 \\
  &  &  &  & 500 & 23.6  \\
  &  &  &  & 700 & 6.94   \\
  \hline \hline
  
\multirow{8}{*}{II} & \multirow{8}{*}{0.2} & \multirow{2}{*}{150} & 10 & - & - \\
  \cline{4-6}  
  &  &  & 50 & - & - \\
  \cline{3-6}
  &  & \multirow{6}{*}{250} & \multirow{5}{*}{10} & 300 & 56.8 \\
  &  &  &  & 400 & 60.7 \\
  &  &  &  & 500 & 89.4  \\
  &  &  &  & 700 & 56.3   \\
  &  &  &  & 1000 & 17.7   \\
  \cline{4-6}  
  &  &  & 50 & -  & - \\  
  \hline \hline
  
 \multirow{8}{*}{III} & \multirow{8}{*}{0.75} & \multirow{3}{*}{150} & \multirow{2}{*}{10} & 700 & 62.8 \\
  &  &  &  & 900 & 16.4 \\
  \cline{4-6}
  &  &  & 50 & - &  -    \\
  \cline{3-6}
  &  &  \multirow{5}{*}{250} & 10 & 300 & 74.6 \\
  \cline{4-6}
  &  &  & \multirow{4}{*}{50} & 500 & 93.6 \\
  &  &  &  & 600 & 78.8 \\
  &  &  &  & 700 & 58.7  \\
  &  &  &  & 800 & 41.9   \\
\hline  
\end{tabular}
\end{center}
\caption{Significances at the LHC at $\sqrt{s}=14$ TeV and 3000 fb$^{-1}$ of integrated luminosity for $m_{\tilde{l}} < M_{\neu{2}}$. Benchmark points with different $M_2/ \mu$ values, belonging to different regions as defined in equations (\ref{higgsino}), (\ref{wino}), (\ref{wino-higgsino}) with Bino-type LSP, are presented which satisfy \gmu \, requirement and are also not excluded by 8 TeV LHC results. $\Delta m_1 = m_{\tilde{l}} - M_{\neu{1}}$ is fixed at 25 GeV for all points. See text for details. }
\label{slltneu2Reach}
\end{table}

This scenario provides a clean signal at the LHC to probe electroweakinos and consequently the strongest bound on electroweakino masses are derived ~\cite{ATLASS2lep,ATLASS3lep,CMSSlep1} for this case. Due to the absence of any signal in LHC Run-I, both ATLAS and CMS exclude $M_{\neu{2}} \sim 730$ GeV for $M_{\neu{1}} \sim 0-350$ GeV, with $m_{\tilde{l}} = (M_{\neu{1}}+M_{\neu{2}})/2$. However, interpreting these exclusion limits for realistic BPs involve a degree of complexity owing to the interplay between three masses ($m_{\tilde{l}},M_{\neu{1}},M_{\neu{2}}$). The relative mass differences $\Delta m_1= m_{\tilde{l}} - M_{\neu{1}}$ and $\Delta m_2= M_{\neu{2}} - m_{\tilde{l}}$ determine the $p_T$ of leptons in the final state which, in turn, dictates the detection efficiency for a particular BP.  Besides, similar to Case (i), CMS and ATLAS use wino-type $\neu{2}$ for there estimation of these bounds. For higher masses of $\neu{2}$, \gmu \, requires smaller $\tilde{l}$ mass, closer to $M_{\neu{1}}$. For $m_{\tilde{l}} = 0.95 M_{\neu{1}}+0.05 M_{\neu{2}}$, CMS sets an upper bound of  $M_{\neu{2}} \sim 730$ GeV as well, but for $M_{\neu{1}} \sim 0-240$ GeV with   $\neu{2}$ decaying to sleptons and leptons democratically. We have used \CheckMATE~\cite{CheckMATE} to estimate bounds for higgsino like $\neu{2}$, for $m_{\tilde{l}} < M_{\neu{2}}$, adopting the results from Ref.~\cite{CMSSlep1}. 

We have chosen two sets of benchmark scenarios for this case. For the first set, $M_{\neu{1}}$ and $m_{\tilde{l}}$ are set to 150 GeV and 175 GeV respectively, while for the second set the corresponding masses are 250 GeV and 275 GeV. Although smaller $\Delta m_1$ values are allowed by \gmu \, due to the presence of soft leptons in these compressed scenarios the DY processes become less efficient. One needs to make use of monojet or dijets to boost the system for these BPs. We have not explored these compressed scenarios in this paper, but invite the interested reader to consult Ref.~\cite{VBF-sl,Monojet-Sl,Barr:2015eva} where $\Delta m_1 \sim 5-25$ GeV has been probed. We have adopted the $3l + \met$ final state, arising from the decays $\neu{2} \rightarrow \tilde{l}/\tilde{l}^* l^{\pm} \rightarrow \neu{1} l^{\mp} l^{\pm}$ and $\chp{1} \rightarrow \tilde{\nu}_{l_L} l^+ \rightarrow \neu{1} \nu_{l_L} l^+$, to probe these scenarios. They can also be probed by same-sign dilepton channel in the case where one lepton is unidentified. However, we have only considered $3 \, l + \met $ channel for this study. The traditional search strategy in $3 \,l + \met$ channel~\cite{CMSSlep1} has been adapted, with optimized cuts for $\sqrt{s} = 14$ TeV. Additionally we have imposed stringent $p_{T_{l_1}} > 30 -100$ GeV cut on the leading lepton ($l_1$), optimized for each BP to maximize the significance. This cut is found to be the strongest discriminant together with large $\met$. The details of the search strategy and efficacy of each cut on the signal and background are shown in Appendix~\ref{SS4}.

Having set the framework of this analysis let us discuss the results. For the BP $(M_{\neu{1}},m_{\tilde{l}}) = (150,175)$ GeV, we derive an exclusion limit of $M_{\neu{2}} \sim 300$ GeV for higgsino-like $\neu{2}$ from the 8 TeV results of the LHC. For $\tan \beta = 10$, \gmu \, is satisfied for $M_{\neu{2}} \sim 1200$ GeV for this BP but due to relatively small production cross-section of higgsinos, we are able to investigate only a fraction of this mass range at the 14 TeV LHC. At $95\%$ CL the exclusion limit $M_{\neu{2}} \sim 850$ GeV can be set with 3000 fb$^{-1}$ integrated luminosity. The details of significance for different masses of $\neu{2}$ are shown in Table~\ref{slltneu2Reach}. For $\tan \beta = 50$, on the other hand, we did not find any point that can explain the \gmu \, excess for the combination of $M_{\neu{1}},m_{\tilde{l}}$ under discussion.

Similarly for the BP $(M_{\neu{1}},m_{\tilde{l}}) = (250,275)$ GeV \gmu \, is not satisfied for any value of $M_{\neu{2}}$ with $\tan \beta = 10$, but $\tan \beta = 50$ allows 
$M_{\neu{2}} \sim 1200$ GeV. The LHC has not set any exclusion limit for these points so far, but at 14 TeV we shall be able to probe upto $M_{\neu{2}} \sim 850$ GeV at $95 \%$ CL.

We should remind the reader that the significances presented in Table~\ref{slltneu2Reach} are dependent on the relative magnitudes of $\Delta m_1$ and $\Delta m_2$, for smaller values of $\Delta m = M_{\neu{2}} - M_{\neu{1}}$. However, the impact is not significant and one such study is presented in Appendix~\ref{DeltaM} for $\Delta m = 50$ GeV. In this context we further add that for $\tan \beta =50$ and $M_{\neu{1}}=150$ GeV, $M_{\neu{2}} \sim 520$ GeV and $\sim 720$ GeV will satisfy $\gmu$ with $m_{\tilde{l}}\approx M_{\neu{2}}$ and $m_{\tilde{l}} = (M_{\neu{1}} + M_{\neu{2}})/2$ respectively. These BPs have better prospects of detection at the LHC, compared to the BPs discussed in previous paragraphs, due to their greater $\met$ acceptance.

\subsubsection{Higgsino LSP}

 For the pure higgsino-like LSP case, Ref.~\cite{Baer:2014kya} has shown that the LHC can probe higgsino-type LSP upto $M_{\neu{2}} \sim 250$ GeV with 1000 fb$^{-1}$ of integrated luminosity in 2 $l$ + 1 $j$ + $\met$ channel. Interestingly if non-thermal DM scenarios are considered, then from Figure \ref{fig-indirect} we have seen that $M_{\neu{1}} \sim 275$ GeV will be excluded by the Fermi-LAT indirect detection experiment.  
 Extrapolating from the significance plot presented in Figure 4 of the Ref.~\cite{Baer:2014kya}, we find that with 3000 fb$^{-1}$ of integrated luminosity the $95\%$ CL exclusion reach can be extended upto $M_{\neu{1}} \approx M_{\neu{2}} \sim 320$ GeV. However for $\tan \beta = 50$, there exist solutions with $M_{\neu{2}} > 320$ GeV which will not be able to be probed by this strategy.  Pure monojet searches also do not work for higgsino LSP~\cite{Baer:Puremono} due to very small $S/B$ ratio.
 
 For example, if $M_1$ is set to be heavy ($\sim$ TeV) $\gmu$ will be satisfied by higgsino like LSP of mass $\sim 400$ GeV and $\sim 500$ GeV for $m_{\tilde{l}} > M_{\neu{3}}$ and $m_{\tilde{l}} < M_{\neu{3}}$ cases, respectively, for $\tan \beta = 50$ and $M_2/\mu =2 $. These points then can be probed by searching for wino-like $\neu{3}$ and $\chpm{2}$. Two such representative points are shown in Table~\ref{higgsinoLSP}. Search strategies described for bino-like LSP are also employed here. Clearly the $m_{\tilde{l}} < M_{\neu{3}}$ will be possible to probe at the LHC, while the $m_{\tilde{l}} > M_{\neu{3}}$  point will remain beyond it's reach. 

\begin{table}[t]
\begin{center}
\begin{tabular}{|c | c |c | c | c | c |} 
\hline
\hline 
  Case & $M_2/\mu$ & $M_{\neu{1}}$ & $\tan \beta$ & $M_{\neu{3}}$ &  Significance($\sigma$) \\ 

         &    & [GeV] &   & [GeV]  &  $\dfrac{S}{\sqrt{(S+B)}}$ \\
\hline

  $m_{\tilde{l}} > M_{\neu{3}}$ & \multirow{2}{*}{2} & 350 & \multirow{2}{*}{50} & 740 & 1.81 \\
  $m_{\tilde{l}} < M_{\neu{3}}$ &  & 450 &  &  937 & 8.61 \\
  
\hline  
\end{tabular}
\end{center}
\caption{Significances at the LHC at $\sqrt{s}=14$ TeV and 3000 fb$^{-1}$ of integrated luminosity for higgsino-like LSP points. For $m_{\tilde{l}} < M_{\neu{3}}$ BPs a mass-gap of $\Delta m_1 = m_{\tilde{l}} - M_{\neu{1}} = 25$ GeV has been maintained. See text for details. }
\label{higgsinoLSP}
\end{table}

\subsection{\label{II}Region - II \, ($M_2/\mu\leq 0.2$)}

For this region $\neu{1}$ is pure bino-type and $\neu{2}$ is pure wino-type if $M_1 \ll M_2 $, and vice versa if $M_2 \ll M_1 $. In contrast the neutralinos will acquire both bino and wino components if $M_1$ and $M_2$ are comparable.  

\begin{figure}[!ht]
\centering
\includegraphics[scale=.4]{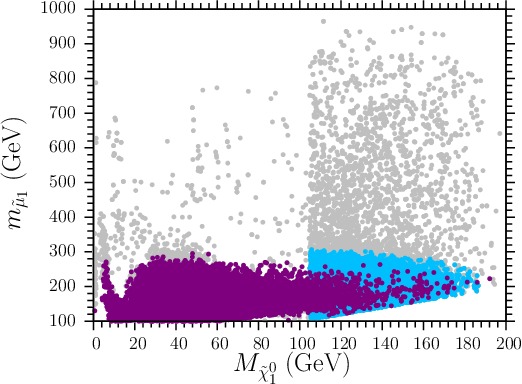}
\includegraphics[scale=.4]{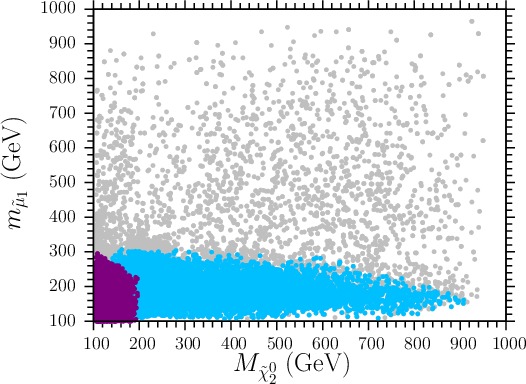}
\includegraphics[scale=.4]{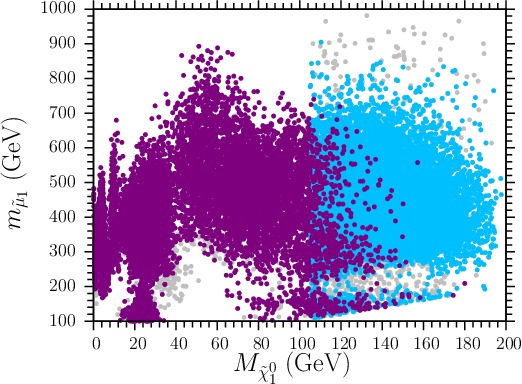}
\includegraphics[scale=.4]{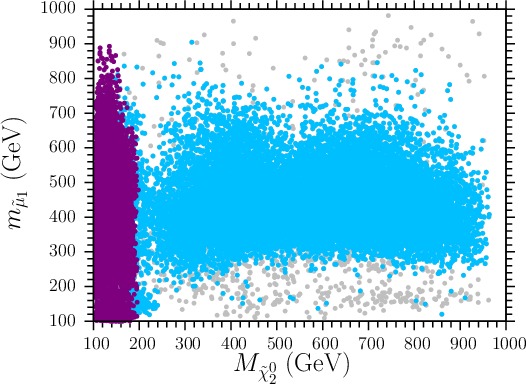}
\caption{Plots in the $m_{\tilde{\mu}_1}-M_{\tilde{\chi}^{0}_{1}}$ and $m_{\tilde{\mu}_1}-M_{\tilde{\chi}^{0}_{2}}$ planes for $\tan\beta=10$ (\textit{upper panel}) and $\tan \beta=50$ (\textit{lower panel}). All points in these plots satisfy the definition given in equation (\ref{wino}) for Region-II. Light gray points in this plot satisfy the constraints given in equation (\ref{mass-constraints}). Light blue points are subset of the gray, satisfy $\gmu$ in the $2 \sigma$ range and $M_2/M_1<1$. Purple points are subset of the gray, satisfy $\gmu$ in the $2 \sigma$ range and also $M_2/M_1>1$.}
\label{fig:wino1}
\end{figure}

In Figure \ref{fig:wino1} we display our results for Region-II in the same planes as in Figure \ref{fig:higgsino1}. The points in these plots all satisfy the definition for Region-II presented in equation (\ref{wino}). \textit{Light gray} points in these plots satisfy the constraints given in equation (\ref{mass-constraints}). \textit{Light blue} are subset of the \textit{gray}, satisfy $\gmu$ in the $2 \sigma$ range and $M_2/M_1<1$. \textit{Purple} points are subset of the \textit{gray}, satisfy $\gmu$ in the $2 \sigma$ range and also $M_2/M_1>1$. For this case $\tilde{\mu}_1$ has to be lighter than $\sim$ 300 GeV for $\tan \beta = 10$. For the \textit{light blue} points, the $\neu{1}$ will essentially be a pure wino, whereas for the \textit{purple} points the $\neu{1}$ can have a large bino component. 
The mass of the  $\tilde{\chi}^{0}_{2}$ however has to be less than 200 GeV, (again for $\tan \beta = 10$) which is an artefact of limiting our scan to $M_1,M_2,\mu < 1$ TeV. Similar to Region-I, a wider range of parameter space satisfies the \gmu \, requirement with $\tan \beta = 50$. 

 This region is also divided into two sub-regions depending upon the nature of the LSP. We should point out that the search strategies discussed for Region-I are also used for this Region. $M_2/\mu$ has been set equal to 0.2 for following analyses.

\subsubsection{Bino LSP}

\subsubsection*{$\mathbf{Case \, (i): m_{\tilde{l}} > M_{\neu{2}}}$}
 
  Among the various neutralinos, the wino is the one that is most abundantly produced at the LHC via s-channel $W$ exchange, due to it's large coupling to the $W$ boson.
The LHC bounds coming from the 8 TeV data are similar to those discussed for the same scenario in Region-I since they were derived by CMS and ATLAS for wino. 
Following the classification mentioned in Region-I we discuss the results for $\Delta m=50$ GeV and $\Delta m \geq m_Z$ cases below.

{\bf $\Delta m=50$ GeV -} This case is of particular interest for wino type $\neu{2}$. We should recall that we are probing these moderately compressed points in the boosted $3 l + 1j + \met $ final state. The dominant production channel for this final state is, $p p \rightarrow \neu{2} \chpm{1} j$. The $\neu{2}$ can decay into $\neu{1}$ accompanied by either an off-shell $\tilde{l_L}$ or $Z$ boson, which in turn decays to yield two leptons. Similarly, $\chpm{1}$ can decay into $\neu{1}$ along with either an off-shell $\tilde{\nu}_{l_L}$ or $W^{\pm}$ boson. If the left-handed sleptons/sneutrinos are not too heavy compared to $\neu{2}/\chpm{1}$, the former dominates over the latter due to unsuppressed $SU(2)_L$ coupling of wino, resulting in sharp enhancement in BR for $\neu{2} \rightarrow \neu{1} l^{\pm} l^{\mp}$ and $\chp{1} \rightarrow \neu{1} l^{+} \nu_l$. Consequently, the aforementioned BP for the wino case with $\tan \beta = 10$, can easily be probed with $> 100 \sigma$ at LHC14. No such wino-like $\neu{2}$ point is allowed for $\tan \beta = 50$.

{\bf $\Delta m \geq m_Z$ -}  A wino-like $\neu{2}$ is consistent with \gmu \, excess upto $M_{\neu{2}} \sim 300$ GeV for $\tan \beta = 10$, and between $300-550$ GeV for $\tan \beta = 50$ with $m_{\tilde{l}} > M_{\neu{2}}$. With 3000 fb$^{-1}$ integrated luminosity the LHC will be able to exclude $M_{\neu{2}} \sim 650$ GeV at $95\%$ CL. In contrast to the simplified case  $M_{\neu{1}}=0$, where the LHC will be able to probe upto $M_{\neu{2}} \approx 500$ GeV at 95$\%$ CL, as demonstrated in Ref.~\cite{Padhi}. The detailed statistical significances of these BPs are shown in Table~\ref{slgtneu2Reach}. A rather low significance is observed for the BP ($M_{\neu{1}},M_{\neu{2}}$) = (150,300) GeV. This is due to the fact that the asymmetric $M_{T_2}$ cut used in the $3 \, l \, + \, \met$ channel is incapable of distinguishing between the signal and the $WZ$ background for $\Delta m \lesssim 200$ GeV. In contrast for Region-I the presence of relatively light $\neu{3}$ and large production cross-section of heavy wino-type $\neu{4}$ compensates for this inefficiency and further improves the efficiency of $\met$ and $\Delta \phi$ cuts as well. Moreover the absence of light $\neu{3}$ makes the signal in the $4 \, l \, + \, \met$ channel non-existent. Consequently the  $95\%$ CL exclusion reach for wino-type $\neu{2}$ is considerably lower than higgsino-type case.

\subsubsection*{$\mathbf{Case \, (ii): m_{\tilde{l}} < M_{\neu{2}}}$} 

We use the same set of benchmark scenarios as discussed in the corresponding case in Region-I. For the BP $(M_{\neu{1}},m_{\tilde{l}}) = (150,175)$ GeV with $\tan \beta = 10$, the \gmu \, excess can be explained by $M_{\neu{2}} \lesssim 300$ GeV, which has been excluded by the LHC. In contrast, for the benchmark $(M_{\neu{1}},m_{\tilde{l}}) = (250,275)$ GeV, \gmu \, will allow $M_{\neu{2}} \sim 1200$ GeV. Similar to higgsino-like $\neu{2}$, the 8 TeV LHC data does not yield any exclusion limit for this BP. Nonetheless, in the upcoming 14 TeV run of the LHC we should be able to set an exclusion limit of $M_{\neu{2}} \sim 1300$ GeV at 95$\%$ CL with 3000 fb$^{-1}$ integrated luminosity. The case $\tan \beta = 50$ does not satisfy the \gmu \, requirement for either of these benchmark scenarios. These results are tabulated in Table~\ref{slltneu2Reach}. With $m_{\tilde{l}} \approx M_{\neu{2}}$ and $m_{\tilde{l}} = (M_{\neu{1}}+M_{\neu{2}})/2$, $M_{\neu{2}} \sim 590$ GeV and $\sim 1200$ GeV will satisfy $\gmu$ for $\tan \beta = 50$ and $M_{\neu{1}}=150$ GeV. Similar to Region-I, these points will have greater possibility of detection at the LHC due to the presence of significant $\met$ in the system, compared to $m_{\tilde{l}}=175$ GeV BPs.

\subsubsection{Wino LSP}

\begin{figure}[!t]
\centering

\includegraphics[scale=.4]{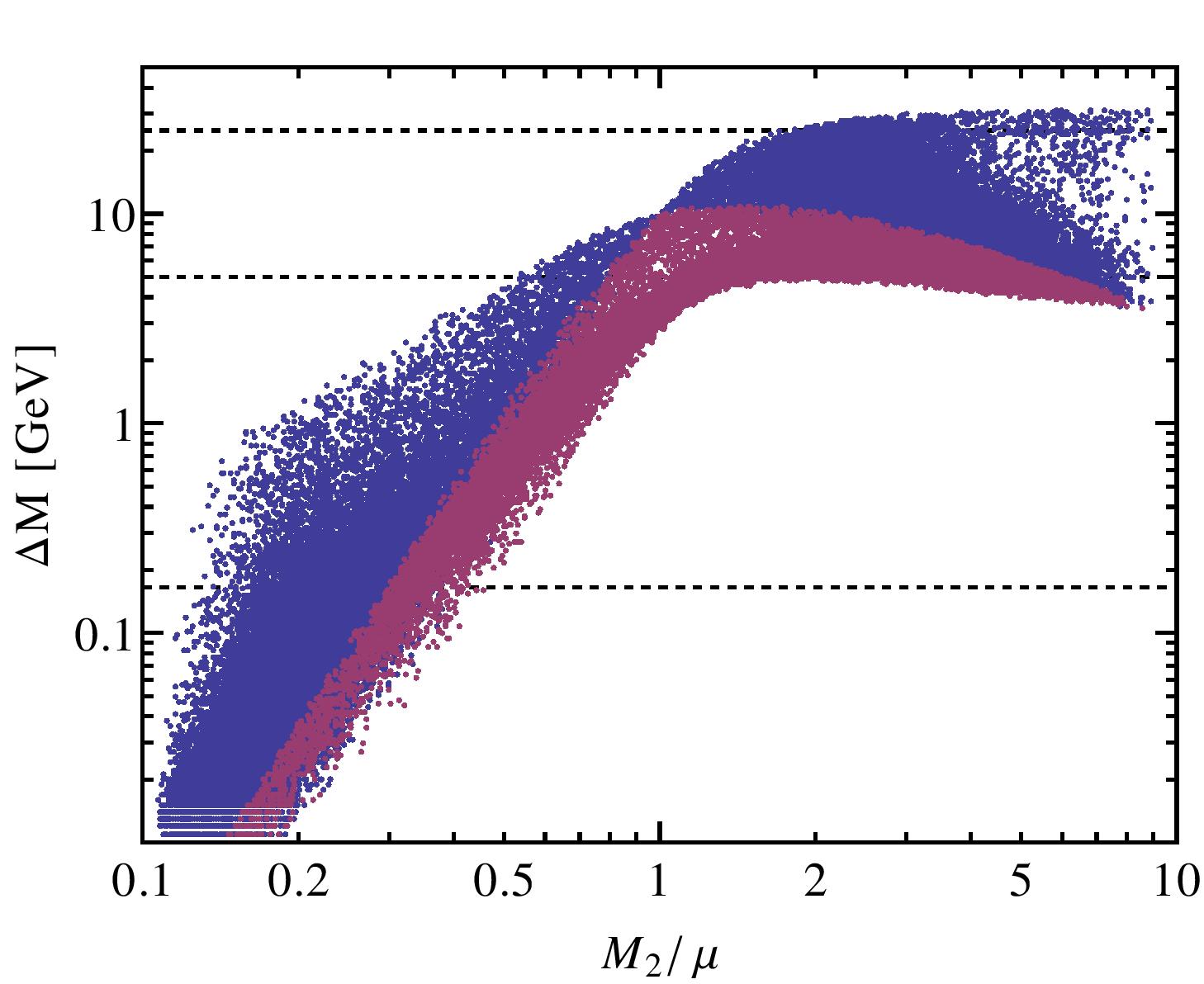}
\caption{$\Delta M = M_{\chpm{1}} - M_{\neu{1}}$ as a function of $M_2/\mu$ for $\tan \beta=10$. {\it Blue} points represent $ min(M_2,\mu) < M_1 < max(M_2,\mu)$, while {\it Purple} points are for $M_1 > max(M_2,\mu)$. The gridlines are drawn for $\Delta M = $ 165 MeV, 5 GeV and 25 GeV respectively.}
\label{wino-chargino}
\end{figure}

  Pure wino-like LSP scenario is already highly constrained from LHC Run-I. For these scenarios the lightest chargino is expected to be degenerate with LSP with a mass-splitting ($\Delta M$) of $\mathcal{O}$(100) MEV~\cite{wino-ch,wino-ch-2loop}. Consequently resulting in unique collider signatures of either disappeared tracks/displaced vertices or long-lived charged particles that do not decay within the detector depending on whether $\Delta M$ is greater or less than $m_{\pi^{\pm}} \sim 140$ MeV. 

   The LHC experiments are performing dedicated searches in both these channels. In disappearing track search strongest bound of $M_{\chpm{1}} \sim 500$ GeV with $\Delta M = 140$ MeV, is presented by the CMS experiment~\cite{CMS:dis-tr}. In contrast for the long-lived charged particle search the strongest bound comes from the ATLAS experiment ($M_{\chpm{1}} \sim 620$ GeV with $\Delta M < 140$ MeV)~\cite{ATLAS:llcp}. The MSSM particle spectra consistent with $\gmu$ can offer both these scenarios~\cite{wino-ch}. In an extreme case when both $\mu$ and SSB sfermion masses are heavy [$\mathcal{O}$(TeV)], $\Delta M$ saturates at $\sim 165$ MeV at 2-loop level~\cite{wino-ch-2loop}. In that case the disappearing track exclusion limit relaxes to $\sim 250$ GeV~\cite{CMS:dis-tr}. Ref.~\cite{Cirelli:wino} has estimated the prospect of this particular scenario at LHC14 and their conservative $95\%$ CL exclusion reach is $\sim 500$ GeV at 3000 fb$^{-1}$ of integrated luminosity.

  Our choice of  $M_{2}/\mu \leq 0.2$ for this region ensures $\Delta M < 165$ MeV when bino is heavy. However when bino is light,  a small mixture of bino in the LSP composition will increase the $\Delta M$ to $\mathcal{O}$(GeV) and the efficacy of above searches will be lost. To illustrate these mass-gaps $\Delta M$ is plotted as a function of $M_2/\mu$ in Figure~\ref{wino-chargino} for $\tan \beta = 10$. The corresponding plot for $\tan \beta = 50$ is similar. In Figure~\ref{wino-chargino}  {\it Blue} points represent $ min(M_2,\mu) < M_1 < max(M_2,\mu)$ scenarios, while {\it Purple} points are for $M_1 > max(M_2,\mu)$. Monojet and VBF searches offer the best possibility to probe those cases. The estimated  monojet $95\%$ CL exclusion reach at 14 TeV LHC run with 3000 fb$^{-1}$ of integrated luminosity is $\sim 400$ GeV~\cite{Cirelli:wino}. However as shown by~\cite{Baer:Puremono} that monojet searches suffer from low $S/B$ ratio and can be dominated by systematic errors. If $5 \%$ systematic error is taken into account then Ref.~\cite{Cirelli:wino} predicts the wino exclusion reach to be $\sim 200$ GeV. In contrast VBF searches don't suffer from low $S/B$ ratio~\cite{VBF-ewk,VBF-sl}. The VBF search performed by Ref.~\cite{VBF-DM} predicts the LHC to probe wino LSP upto $\sim 600$ GeV at 1000 fb$^{-1}$ of integrated luminosity (However see~\cite{Cirelli:wino,Berlin:VBF}). Again a pure wino-like LSP scenario is severely constrained ($\sim 575$ GeV) by the new results from Fermi-LAT if non-thermal DM scenarios are considered.
  
  Considering smuons are not much heavier than $\neu{1}$, the $\gmu$ excess can be explained ny by wino-like LSP of mass $\sim 500$ GeV for $\tan \beta = 50$. From the above discussion it is evident that most of the wino-like LSP scenarios are either already ruled out by Run-I or will be excluded in the upcoming run of the LHC.
  


\subsection{\label{III}Region - III \, ($ 0.2 < M_2/\mu < 2$)}

This Region is characterised by comparable values of $M_2$ and $\mu$ and neutralinos will have both wino and higgsino components. In addition, they may contain a large bino component as well depending on the relative magnitude of $M_{1}$ in comparison with $M_2$ and $\mu$.  

\begin{figure}[!ht]
\centering
\includegraphics[scale=.4]{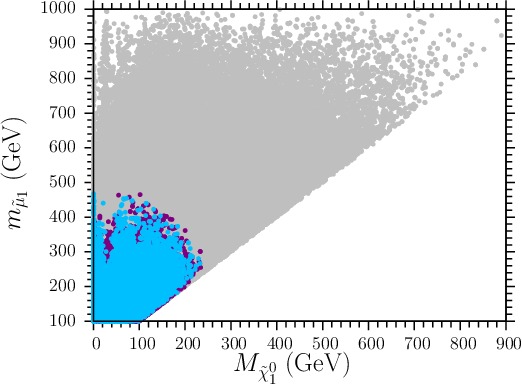}
\includegraphics[scale=.4]{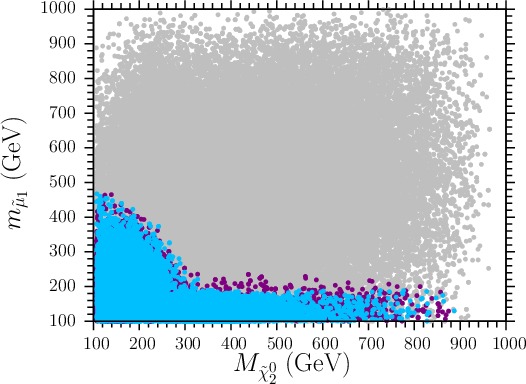}
\includegraphics[scale=.4]{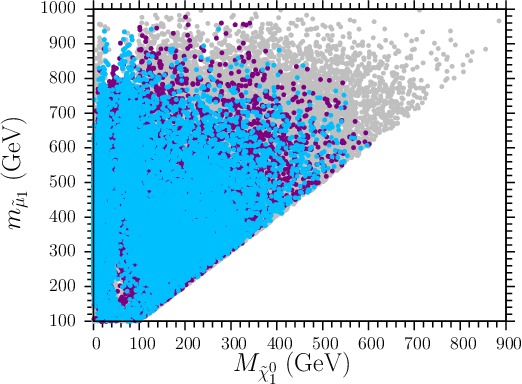}
\includegraphics[scale=.4]{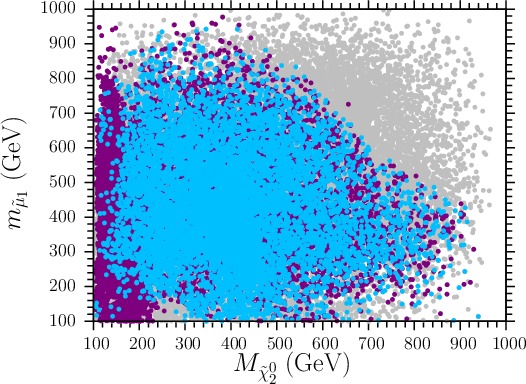}
\caption{Plots in the $m_{\tilde{\mu}_1}-M_{\tilde{\chi}^{0}_{1}}$ and $m_{\tilde{\mu}_1}-M_{\tilde{\chi}^{0}_{2}}$ planes for $\tan\beta=10$ (\textit{upper panel}) and $\tan \beta=50$ (\textit{lower panel}). All points in these plots satisfy the definition given in equation (\ref{wino-higgsino}) for Region-III. Light gray points in this plot satisfy the constraints given in equation (\ref{mass-constraints}). As before, the light blue and purple points satisfy $\gmu$ in the $2 \sigma$ range. Light blue points are subset of the gray,  and also satisfy $M_1 < \mu < M_2$. Purple points are subset of the gray,  and also satisfy $M_1 < M_2 < \mu$.}
\label{fig:wh1}
\end{figure}

\begin{figure}[!ht]
\centering
\includegraphics[scale=.4]{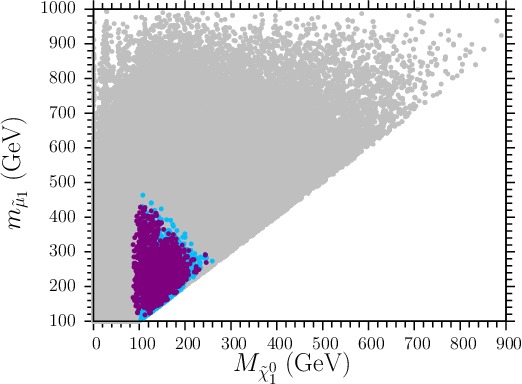}
\includegraphics[scale=.4]{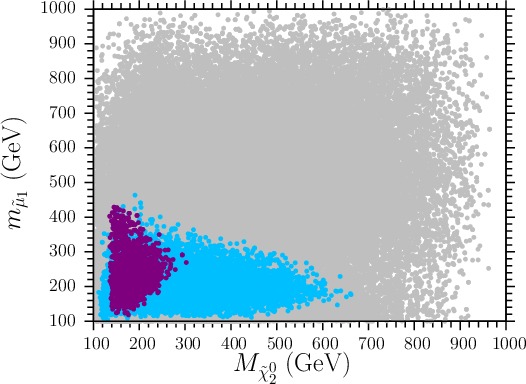}
\includegraphics[scale=.4]{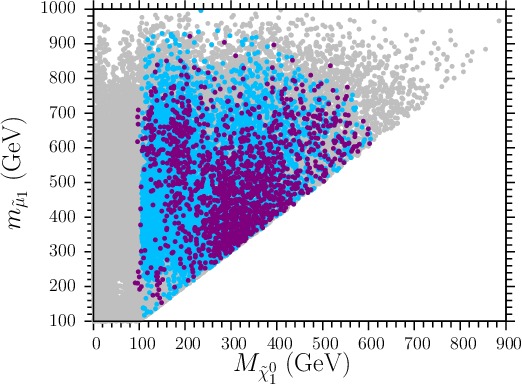}
\includegraphics[scale=.4]{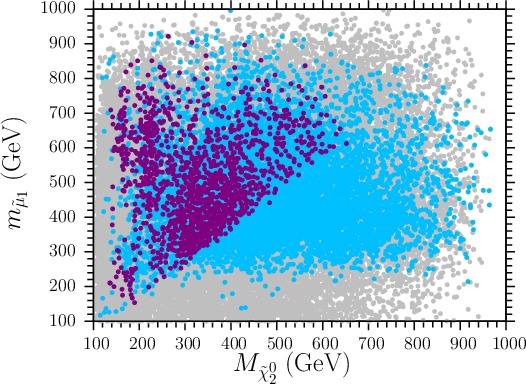}
\caption{Plots in the $m_{\tilde{\mu}_1}-M_{\tilde{\chi}^{0}_{1}}$ and $m_{\tilde{\mu}_1}-M_{\tilde{\chi}^{0}_{2}}$ planes for $\tan\beta=10$ (\textit{upper panel}) and $\tan \beta=50$ (\textit{lower panel}). Light gray points in this plot satisfy the constraints given in equation (\ref{mass-constraints}). As before, the light blue and purple points satisfy $\gmu$ in the $2 \sigma$ range. Light blue points are subset of the gray,  and also satisfy $M_2/M_1<1$ and $M_1/\mu<1$. Purple points are subset of the gray,  and also satisfy $M_2/M_1>1$ and $M_1/\mu>1$.}
\label{fig:wh2}
\end{figure}

 In Figure \ref{fig:wh1} and \ref{fig:wh2} we display our results for Region-III in the same planes as in Figure \ref{fig:higgsino1}. All points in these plots satisfy the definition for Region-III given in equation (\ref{wino-higgsino}). \textit{Light gray} points in these plots satisfy the constraints given in equation (\ref{mass-constraints}). As before, the \textit{light blue} and \textit{purple} points satisfy $\gmu$ in the $2 \sigma$ range. In Figure \ref{fig:wh1}, the \textit{light blue} points are subset of the gray, and satisfy $M_1 < \mu < M_2$. \textit{Purple} points are subset of the \textit{gray}, and satisfy $M_1 < M_2 < \mu$. On the other hand, in Figure \ref{fig:wh2}, the \textit{light blue} points satisfy $M_2/M_1<1$ and $M_1/\mu<1$ and, \textit{purple} points satisfy $M_2/M_1>1$ and $M_1/\mu>1$. For both cases $\tilde{\mu}_1$ has to be lighter than $\sim$ 450 GeV for $\tan \beta = 10$. In Figure \ref{fig:wh1}, for both \textit{light blue} and \textit{purple} points, the $\neu{1}$ will have sizable wino and higgsino components. For this region the parameter space available for \textit{light blue} and \textit{purple} points are almost identical.  In contrast, in Figure \ref{fig:wh2} the \textit{light blue} points represent a $\neu{1}$ with a sizable wino and bino component, whereas for \textit{purple} points the bino and higgsino components can be substantial. For the \textit{purple} points the $\tilde{\chi}^{0}_{1}$ and $\tilde{\chi}^{0}_{2}$ masses are bounded as follows: 
100 GeV $\lesssim M_{\tilde{\chi}^{0}_{1}} \lesssim $ 250 GeV and 
140 GeV $\lesssim M_{\tilde{\chi}^{0}_{2}} \lesssim $ 300 GeV for $\tan \beta =10$. A considerably larger parameter space is allowed for $\tan \beta =50$ (for both figures). 


\subsubsection{Bino LSP}

We have set $M_2/\mu=0.75$ for the BPs analyzed in this sub-section.

\subsubsection*{$\mathbf{Case \, (i): m_{\tilde{l}} > M_{\neu{2}}}$}

Electroweakinos belonging to this region will also be sufficiently produced at the LHC. Although the production cross-section will be smaller than the pure wino-type $\neu{2}$ case, it is compensated by the presence of lighter $\neu{3}$ and $\neu{4}$, resulting in contributions from some secondary channels in addition to the dominant $\neu{2} \chpm{1} j$ channel. The LHC bounds, BPs and search strategies adopted for $m_{\tilde{l}} > M_{\neu{2}}$ case is similar to Region-I. 

{\bf $\Delta m=50$ GeV -} Analogous to pure wino and higgsino cases, this scenario arises only in the wino-higgsino case for $\tan \beta = 10$. This BP [($M_{\neu{1}}, M_{\neu{2}}$)= (150, 200) GeV] should be readily accessible at LHC 14 ($> 50 \sigma$).

{\bf $\Delta m \geq m_Z$ -}  With $M_{\neu{1}}=150$ GeV, the \gmu \, excess can be explained by $M_{\neu{2}}$ upto $\sim 300$ GeV for $\tan \beta = 10$, and between 300-650 GeV for $\tan \beta =50$. The presence of light 
$\neu{3,4}$ for the BPs belonging to this region, makes the prospect of 
excluding ($\sim 975$ GeV) both $\tan \beta =10$ and $\tan \beta = 50$ points at $95\%$ CL, encouraging in the upcoming high luminosity run of the LHC. The statistical significances for all BPs are demonstrated in Table~\ref{slgtneu2Reach}. The corresponding mass-reach for this scenario with $M_{\neu{1}}=0$ GeV, quoted by Ref.~\cite{Padhi}, is 700 GeV at 95$\%$ CL, but for $M_2 \approx \mu$.

\subsubsection*{$\mathbf{Case \, (i): m_{\tilde{l}} < M_{\neu{2}}}$} 

With $m_{\tilde{l}} < M_{\neu{2}}$ the \gmu \, requirement is satisfied by a wide range of masses and $\tan \beta$ values for the wino-higgsino case. For the BP $(M_{\neu{1}},m_{\tilde{l}}) = (150,175)$ GeV, \gmu \, allows $M_{\neu{2}} \sim 900$ GeV for $\tan \beta = 10$ but no point is allowed for $\tan \beta = 50$. We derived the 8 TeV LHC exclusion bounds for this scenario to be $\gtrsim 500$ GeV. In the high luminosity (3000 fb$^{-1}$) run of the LHC , one should be able to extend the exclusion limit upto $\sim 1350$ GeV at 95$\%$ CL. For the BP $(M_{\neu{1}},m_{\tilde{l}}) = (250,275)$ GeV, \gmu is explained by both $\tan \beta$ values. While $\tan \beta = 10$ allows $M_{\neu{2}} \sim 300$ GeV (which is again easily accessible at LHC 14), $\tan \beta = 50 $ allows $M_{\neu{2}} \sim 500-800$ GeV. Similar to previous sections no bound on $\neu{2}$ mass is offered by LHC 8 data for this BP. The extended exclusion limit at 95$\%$ CL will be similar to the previous BP. The statistical significance for BPs are shown in Table~\ref{slltneu2Reach}.  For $m_{\tilde{l}} \approx M_{\neu{2}}$ and $m_{\tilde{l}} = (M_{\neu{1}}+M_{\neu{2}})/2$, $M_{\neu{2}} \sim 620$ GeV and $\sim 875$ GeV will satisfy $\gmu$ with $\tan \beta = 50$ and $M_{\neu{1}}=150$ GeV.\\

\subsubsection{Wino-Higgsino LSP}

\begin{table}[t]
\begin{center}
\begin{tabular}{|c | c |c | c | c | c| c|} 
\hline
\hline 
  $M_2/\mu$ &   &   $M_{\neu{1}}$  & $\tan \beta$ & $M_{\neu{2}}$ & $M_{\neu{3}}$ &  Significance($\sigma$) \\ 

         &     & [GeV] &   & [GeV]  & [GeV] &  $\dfrac{S}{\sqrt{(S+B)}}$ \\
\hline

 \multirow{6}{*}{0.75} & \multirow{3}{*}{$m_{\tilde{l}} < M_{\neu{2}}$} & 100 & 10 & 187 & 223 & 24.5 \\  
 \cline{3-7}
          &  & 300 & \multirow{2}{*}{50} & 419 & 440 & 4.56 \\
          &  & 600 &                     & 794 & 802 & 0.29  \\
 
 \cline{2-7}
 
 & \multirow{3}{*}{$m_{\tilde{l}} > M_{\neu{3}}$} & 100 & 10 & 187 & 223 & 8.35 \\
 \cline{3-7}
          &  & 300 & \multirow{2}{*}{50} & 419 & 440 & 2.41 \\
          &  & 500 &                     & 666 & 679 & 0.90  \\ 

\hline                  

\multirow{6}{*}{1} & \multirow{3}{*}{$m_{\tilde{l}} < M_{\neu{2}}$} & 120 & 10 & 182 & 240 & 147.8 \\
\cline{3-7} 
          &  & 350 & \multirow{2}{*}{50} & 404 & 463 & 13.1 \\
          &  & 600 &                     & 624 & 713 & 1.82  \\
 
 \cline{2-7}
 
 & \multirow{3}{*}{$m_{\tilde{l}} > M_{\neu{3}}$} & 100 & 10 & 163 & 220 & 20.92 \\
 \cline{3-7}
          &  & 300 & \multirow{2}{*}{50} & 354 & 413 & 1.19 \\
          &  & 550 &                     & 599 & 661 & 0.25  \\

\hline  
\end{tabular}
\end{center}
\caption{Significances at the LHC at $\sqrt{s}=14$ TeV and 3000 fb$^{-1}$ of integrated luminosity for wino-higgsino like LSP points. For $m_{\tilde{l}} < M_{\neu{2}}$ BPs a mass-gap of $\Delta m_1 = m_{\tilde{l}} - M_{\neu{1}} = 25$ GeV has been maintained.}
\label{whLSP}
\end{table}

 For wino-higgsino LSP we explore two set of BPs with $M_2/\mu$ values 0.75 and 1. Both these cases are characterised by three light neutralinos along with both charginos with $M_{\neu{3}} - M_{\neu{1}} \sim 100-200$ GeV. The essential difference between these two scenarios is, while $\Delta m = M_{\neu{2}} - M_{\neu{1}} \sim 100$ GeV for $M_2/\mu=0.75$, the same is $\sim 50$ GeV for $M_2/\mu=1$. This leads to different collider reaches for these two scenarios owing to  production of on-shell or off-shell $W,Z$, respectively, in $\neu{2}$ decay chain. Both $m_{\tilde{l}} < M_{\neu{2}}$ and $m_{\tilde{l}} > M_{\neu{3}}$ cases have been explored for these BPs. Having set $M_1=1$ TeV, $\gmu$ is satisfied for $M_2/\mu=0.75$ upto $M_{\neu{1}} \sim 500$ GeV and 600 GeV for $m_{\tilde{l}} > M_{\neu{3}}$ and $m_{\tilde{l}} < M_{\neu{2}}$ respectively with $\tan \beta = 50$. Corresponding upper bounds on $M_{\neu{1}}$, for $M_2/\mu=1$ are 550 GeV and 600 GeV.

Monojet searches can probe these points upto $M_{\neu{1}} \sim 275$ GeV and 250 GeV at $95 \%$ CL for $M_2/\mu=0.75$ and 1 respectively with 3000 fb$^{-1}$ of integrated luminosity. However $S/B$ for all these BPs $\lesssim 5 \%$. Hence any source of large systematic error will make monojet search strategy futile for these BPs. Again we can probe these BPs by searching for heavier electroweakinos in multi-lepton + $\met$ channel. For $m_{\tilde{l}} > M_{\neu{3}}$ BPs belonging to both $M_2/\mu$ values, we adopt search strategies discussed in Appendices~\ref{SS2},\ref{SS3},\ref{SS5} and \ref{SS6}. In addition due to $\Delta m$ being $\sim 50$ GeV for $M_2/\mu =1$ BPs, we searched for them by means of search strategy of Appendix~\ref{SS1} as well. In contrast for $m_{\tilde{l}} < M_{\neu{2}}$ BPs the search strategy of Appendix~\ref{SS4} is only used. The $95\%$ CL exclusion limit set for $M_{\neu{1}} \sim 340$ GeV and 390 GeV, respectively, for $m_{\tilde{l}} > M_{\neu{3}}$ and $m_{\tilde{l}} < M_{\neu{2}}$ cases with $M_2/\mu=0.75$. The analogous limits for $M_2/\mu=1$ are 250 GeV and 590 GeV. The expected combined statistical significances at 3000 fb$^{-1}$ of integrated luminosity for these BPs are tabulated in Table~\ref{whLSP}. 

Interestingly the collider reach is higher for $M_2/\mu =1$, compared to $M_2/\mu=0.75$, when $m_{\tilde{l}} < M_{\neu{2}}$ but lower when $m_{\tilde{l}} > M_{\neu{3}}$. This is due to the fact that $\Delta m \sim 50$ GeV for $M_2/\mu=1$ BPs results in loss of sensitivity of search strategies for $m_{\tilde{l}} > M_{\neu{3}}$ scenarios (Appendices~\ref{SS2},\ref{SS3},\ref{SS5} and \ref{SS6}), which requires the presence of on-shell $W$ and/or $Z$. The search strategy of Appendix~\ref{SS1} is only helpful in increasing the sensitivity of $M_{\neu{1}}=100$ GeV point. On the other hand the same $\Delta m \sim 50$ GeV ensures higher BR of $\neu{2} \rightarrow \tilde{l}/\tilde{l}^* l^{\pm}$ decay, while prohibiting $\neu{2} \rightarrow \neu{1} Z/h , \chpm{1}W^{\mp}$ decays for $m_{\tilde{l}} < M_{\neu{2}}$. Hence the increase in efficacy of the search strategy of Appendix~\ref{SS4}.   

 It should be noted that similar to bino-higgsino LSP case, wino-higgsino LSP-nucleon scattering cross-section can also be high. However this can lead to strong direct detection constraints in non-thermal DM scenarios only. For thermal scenarios small relic abundance for wino-higgsino LSP results in suppression of these constraints.

 \subsection{Compressed electroweakino spectra}  Although we have not probed compressed scenarios with mass-splitting, $\Delta m < 50$ GeV but in this subsection we have collected various results available in the literature and extended them in certain cases. It has been mentioned earlier that Ref.~\cite{Baer:2014kya} have predicted that the LHC at 14 TeV will be able to probe higgsino-type LSP upto $\sim 250$ GeV with $\Delta m \sim 10-30$ GeV at 1000 fb$^{-1}$ of integrated luminosity in the monojet+di-lepton+$\met$ channel. Extrapolating from the significance plot presented in Figure 4 of that paper, we find that with 3000 fb$^{-1}$ of integrated luminosity the $95\%$ CL exclusion reach can be extended upto $M_{\neu{1}} \approx M_{\neu{2}} \sim 320$ GeV.

Moreover using the SM backgrounds provided in the same paper we have  set an approximate $95\%$ CL exclusion reach for bino-type LSP with wino-type next-to-lightest SUSY particles (NLSPs) as well. In this case the reach is expected to be $\sim 375$ GeV at 3000 fb$^{-1}$ of integrated luminosity for $\Delta m \sim 10$ GeV. Finally we should  recall that for pure wino-type LSP with $\mathcal{O}$(1 GeV) mass-splitting between $\chpm{1}$ and $\neu{1}$, the corresponding reach is $\sim 400$ GeV~\cite{Cirelli:wino} in monojet analysis.

 \subsection{ Sleptons}     
 
  Finally we conclude our discussion on the role the LHC will play to probe  the \gmu \, parameter space by examining the mass reach of the slepton at 14 TeV. For $m_{\tilde{l}} < M_{\neu{2}}$ the sleptons can be studied at the LHC in non-resonant di-lepton channel by means of their direct production and decay to the LSP. For certain scenarios this channel can provide stronger constraints for the \gmu \, parameter space compared to the constraints from probing electroweakinos. Particularly in scenarios when wino and higgsinos are decoupled and only bino-smuon loop contributes to $\gmu$. Probing the sleptons directly, offers the only possibility to search for these points at colliders. Setting $M_2,\mu \sim 5$ TeV, $\gmu$ is satisfied for $M_{\neu{1}} \approx m_{\tilde{l}} \sim 375$ GeV and 625 GeV for $\tan \beta = 10$ and 50 respectively. From the discussions in the following paragraphs it will be evident that a large portion that parameter space can also be probed at the LHC at 14 TeV by searching for smuons directly. 
  
  Ref.~\cite{Eckel:2014dza} has recently investigated the slepton mass reach for varying nature of the LSP. In contrast we have restricted ourselves to only bino-type LSP in this paper, as argued earlier. For left-handed sleptons with bino-type LSP, Ref.~\cite{Eckel:2014dza} has established a $95\%$ CL exclusion limit of $\lesssim 550$ GeV with $M_{\neu{1}} \sim 0-250$ GeV at 100 fb$^{-1}$ of integrated luminosity with $\Delta m_1 = m_{\tilde{l}} - M_{\neu{1}} \sim 70$ GeV. The corresponding predicted bound for right-handed slepton is $\lesssim 450$ GeV with $M_{\neu{1}} \sim 0-150$ GeV.  However, the definition of statistical significance of Ref.~\cite{Eckel:2014dza}, $\sigma=S/\sqrt{B}$, is different from our definition of $S/\sqrt{S+B}$\footnote{The difference in two definitions of statistical significances, discussed here, is important when $S \gtrsim B$. For $S \ll B$ they yield the same statistical significance.}. To be consistent with the rest of the paper we have extracted the background yield of Ref.~\cite{Eckel:2014dza} and extrapolated the $95\%$ CL exclusion limit for sleptons at 3000 fb$^{-1}$, with our definition of statistical significance.  We found the corresponding limits to be $\sim 775$ GeV for $\tilde{l}_L$ and $\sim 670$ GeV for $\tilde{l}_R$ with $M_{\neu{1}}=100$ GeV. An astute reader can readily notice from Figure 9 of the aforementioned reference, that similar conclusions can be drawn for $M_{\neu{1}}=150$ GeV as well.

  The exclusion limits will be much weaker for compressed scenarios. For 100 fb$^{-1}$ luminosity Ref.~\cite{Monojet-Sl} has shown that the $2 \sigma$ exclusion limits are $m_{\tilde{l}_L} \sim 175-200$ GeV and $m_{\tilde{l}_R} \sim 125-150$ GeV, with $\Delta m_1 \sim 5-20$ GeV. Adapting the same approach described in the previous paragraph we have also extended the results of Ref.~\cite{Monojet-Sl} for 3000 fb$^{-1}$ integrated luminosity. For $\tilde{l}_L$ the $95\%$ CL exclusion limits are $\sim 320 \, (275)$ GeV with $\Delta m_1 \sim 5 \, (20)$ GeV respectively. The corresponding limits for $\tilde{l}_R$ are $\sim 250 \, (225)$ GeV. From the above discussion we note here that no limits are available for slepton masses, to the best of our knowledge, for $\Delta m_1 \sim 20-70$ GeV. 

  Finally if sleptons are extremely degenerate with the LSP, long-lived 
 charged particle searches can be helpful in probing such scenarios. The 8 TeV results of the ATLAS experiment set lower bounds on $m_{\tilde{l}} \sim 385-440$ GeV for $\tan \beta = 10-50$. In a recent analysis Ref.~\cite{Feng:llcp} has predicted the LHC reach for these scenarios, at 14 TeV and 3000 fb$^{-1}$ integrated luminosity, to be  $\sim 1.3$ TeV for $\tilde{l}_L$ and $\sim 1.05$ TeV for  $\tilde{l}_R$ respectively.

\section{Conclusions}\label{conclusions}

\begin{table}[!htp]
\begin{center}
\begin{tabular}{|c |c | c |c | c | c | c | c |} 
\hline
\hline 
 
 $M_{\neu{1}}$ & Region & $M_2/\mu$ &   & \multicolumn{2}{|c|}{$M_{\neu{2}}$ [GeV]} & \multicolumn{2}{|c|}{$m_{\tilde{\l}} \, (< M_{\neu{2}})$ [GeV]} \\
\cline{5-8}  
 [GeV] & & & & $\gmu$ & LHC  & $\gmu$  & LHC  \\

\hline

\multirow{6}{*}{150 -- 250} & \multirow{2}{*}{I} & \multirow{2}{*}{2} &   
$m_{\tilde{l}} > M_{\neu{2}}$ & 200 (550) & 975 & \multirow{2}{*}{270 (520)} & \multirow{3}{*}{775 ($\tilde{l}_L$)} \\
 &  &  &  $m_{\tilde{l}} < M_{\neu{2}}$ &  1200 (720) & 850 &  & \\
 \cline{2-7}
 & \multirow{2}{*}{II} & \multirow{2}{*}{0.2} & $m_{\tilde{l}} > M_{\neu{2}}$ & 300 (550) & 650 & \multirow{2}{*}{310 (670)} & \\
\cline{8-8} 
 &  &  & $m_{\tilde{l}} < M_{\neu{2}}$ &  200 (1200) & 1300 & & \multirow{3}{*}{670 ($\tilde{l}_R$)} \\
 \cline{2-7}
 &  \multirow{2}{*}{III} & \multirow{2}{*}{0.75} & $m_{\tilde{l}} > M_{\neu{2}}$ & 300 (650) & 975 & \multirow{2}{*}{300 (620)} & \\  
 &  &  & $m_{\tilde{l}} < M_{\neu{2}}$ & 900 (875) & 1350 &  & \\
\hline  
\end{tabular}
\end{center}
\caption{Summary of our $\gmu$ scan and subsequent collider analysis with $M_{\neu{1}} \sim 150-250$ GeV (large bino component). $\gmu$ allowed $M_{\neu{2}}$ and $m_{\tilde{l}}$ at $2 \sigma$, have been tabulated for $m_{\tilde{l}} > M_{\neu{2}}$ and $m_{\tilde{l}} < M_{\neu{2}}$ cases for all three regions of parameter space, as defined in equations (\ref{higgsino}), (\ref{wino}), (\ref{wino-higgsino}). In columns 5 and 7 the $\gmu$ allowed values presented are for $\tan\beta=10$. The corresponding $\tan \beta=50$ values are shown within parenthesis. }
\label{Conclusion}
\end{table}

 In this paper we have investigated the weak scale MSSM parameter space, within collider constraints, that explains the BNL measured muon $\gmu$ excess at $2\sigma$ significance level, and we then examined the prospects of probing the parameter space at the future high luminosity run of the LHC. The parameter space scan is performed for two values of $\tan \beta$ (10 and 50). We find that for $\tan \beta = 10$, the $\gmu$ excess can be resolved for relatively smaller masses of $\neu{1}$ ($\lesssim 300$ GeV) and $\tilde{\mu}_1$ ($\lesssim 500$ GeV). The corresponding upper bounds for $\tan \beta = 50$ are $\lesssim 650$ GeV and $\lesssim 1$ TeV respectively. In contrast the upper bound on $M_{\neu{2}}$ is $\sim 1$ TeV for both $\tan \beta$ values. It should be noted that these upper bounds are limited to a degree since we scanned the parameter space upto 1 TeV for $M_1,M_2,\mu,m_{\tilde{\mu}_L}$ and $m_{\tilde{\mu}_R}$.  However our collider study is not restricted to these bounds. We searched for electroweakinos at the LHC, beyond these bounds, whenever necessary, and the relevant discussions are presented in Section~\ref{results-2}. 
 
   We did not impose DM relic abundance or any direct and indirect detection constraint on the parameter space. If non-thermal DM scenarios are considered and indirect detection bounds are taken into consideration, the null results from dwarf galaxies of the Milky Way by Fermi-LAT collaboration will exclude wino-type ($\geq 90 \%$) DM upto $\sim 575$ GeV, and higgsino-type DM upto $\sim 275$ GeV, but it will not impose any constraint on bino-type DM. However these constraints are negligible for thermal wino/higgsino-type DM scenarios owing to a depleted relic abundance. In addition the astrophysical uncertainties are large for indirect detection bounds. The direct detection experiments can also apply strong constraints, especially on bino-higgsino type DM. These bounds can also be relaxed by assuming light $m_A$, but then one needs to consider constraints from Br($B_s \rightarrow \mu^+ \mu^− $) and Br($b \rightarrow s\gamma$). More importantly, direct detection bounds suffer from large uncertainties in proton properties.
 Nevertheless the parameter space we studied can be further constrained if these bounds cannot be evaded.

  We have further divided the parameter space, which satisfy $\gmu$ into three distinct regions based on the relative wino and higgsino content of $\neu{1}$ and $\neu{2}$ as defined in equations (\ref{higgsino}), (\ref{wino}), (\ref{wino-higgsino}). Each of these regions are sub-divided depending on the nature of the LSP. While wino and higgsino LSP scenarios can be probed by searching for the LSP directly along with degenerate $\chpm{1}$ and $\neu{2}$ (higgsino only), for bino LSP all searches for LSP at the LHC will give null result due to its small production rate. Hence for bino-like LSP we have searched for heavier electroweakinos together with sleptons and predicted their $95\%$ CL exclusion limit at 3000 fb$^{-1}$ integrated luminosity for $M_{\neu{1}} = 100-250$ GeV, with $\Delta m = M_{\neu{2}}-M_{\neu{1}} \geq 50$ GeV. The exclusion limits obtained for $m_{\tilde{l}} < M_{\neu{2}} \, (m_{\tilde{l}} > M_{\neu{2}})$ are as follows:  
  \begin{itemize}
  \item Higgsino-type $\neu{2}$ (Region-I):  $\sim 850 \, (975)$ GeV,
  \item Wino-type $\neu{2}$ (Region-II): $\sim 1300 \, (650)$ GeV,
  \item Wino-higgsino type $\neu{2}$ (Region-III): $\sim 1350 \, (975)$ GeV. 
  \end{itemize}  
  On the other hand, extrapolating the results from Ref.~\cite{Eckel:2014dza}, the corresponding limits on the sleptons (degenerate 1st and 2nd generation), with the same range of $\neu{1}$ mass and $\Delta m_1 = m_{\tilde{l}} - M_{\neu{1}} \gtrsim 70$ GeV, are: 
  \begin{itemize}
  \item Left-handed slepton, $\tilde{l}_L$: $\sim 775$ GeV,
  \item Right-handed slepton, $\tilde{l}_R$: $\sim 670$ GeV.
  \end{itemize}
 A summary of these results for $M_{\neu{1}} \sim 150 -- 250$ GeV is presented in Table~\ref{Conclusion} \footnote{The $\gmu$ upper bounds and the LHC reaches shown for slepton masses in the Table~\ref{Conclusion} are for $m_{\tilde{l}} < M_{\neu{2}}$. Otherwise for lighter $ M_{\neu{2}}$, $\gmu$ upper bound on $m_{\tilde{l}_L}$ are $\sim 850-1000$ GeV for 3 regions pertaining to our analysis. In these cases the search strategy for direct production of sleptons, discussed in Ref.~\cite{Eckel:2014dza}, becomes less efficient due to lower BR of $\tilde{l}_L \rightarrow \neu{1} \, l$ decay and $\tilde{l}_L$ decays pre-dominantly to wino-type heavier electroweakinos resulting in cascade decays. $\tilde{l}_R$ decays remains unaffected though. However these points can be easily probed at the LHC by searching for light electroweakino spectra.}. In contrast the corresponding  $95\%$ CL exclusion limits on non-bino like LSP with $m_{\tilde{l}} < M_{\neu{2}} \, (m_{\tilde{l}} > M_{\neu{2}})$ are as follows:
\begin{itemize}
  \item Higgsino-type $\neu{1}$ (Region-I): $> 450$ ($\sim 300$) GeV,
  \item Wino-type $\neu{1}$ (Region-II): $\sim 620$ GeV,
  \item Wino-higgsino type $\neu{1}$ (Region-III - $M_2/\mu=0.75$): $\sim 390 \, (340)$ GeV, 
   \item Wino-higgsino type $\neu{1}$ (Region-III - $M_2/\mu=1$): $\sim 590 \, (250)$ GeV, 
  \end{itemize}  
   A summary of searches and corresponding mass reaches for non-bino type LSP is shown in Table~\ref{Conclusion2}.

\begin{table}[!htp]
\begin{center}
\resizebox{\columnwidth}{!}{%
\begin{tabular}{|c |c | c | c | c | c | c | c|} 
\hline
\hline 
 
  LSP & Region & $M_2/\mu$ &  Conditions  & \multicolumn{2}{|c|}{$M_{\neu{1}}$ [GeV]} & Search & LHC Energy\\
\cline{5-6}  
 type & & & & $\gmu$ & LHC  & Strategy  & and Luminosity  \\

\hline

\multirow{3}{*}{higgsino} & \multirow{3}{*}{I} & $\geq 2$ & -- & 500 & 320 & di-lepton + $\met$ & \multirow{3}{*}{14 TeV, 3000 fb$^{-1}$} \\
\cline{3-7}
  &   & \multirow{2}{*}{2} & $m_{\tilde{l}} > M_{\neu{3}}$ & 400 & $\sim 300 $ & \multirow{2}{*}{multi-lepton + $\met$} & \\
  &   &   & $m_{\tilde{l}} < M_{\neu{3}}$ & 500 & $> 450$ & & \\
  \hline  
                
\multirow{5}{*}{wino} & \multirow{5}{*}{II} & \multirow{5}{*}{$\leq 0.2$} & $\Delta M < 140$ MeV & \multirow{5}{*}{500} & 620 & long lived charged particle & \multirow{3}{*}{8 TeV, 20 fb$^{-1}$} \\
\cline{4-7}
  &  &  & $ \Delta M = 140$ MeV &  & 500 & \multirow{3}{*}{disappearing track} & \\ 
  &  &  & \multirow{2}{*}{$ \Delta M \approx 165$ MeV} &  & 250 &  & \\ 
\cline{8-8}  
  &  &  &  &  & 500 & & \multirow{2}{*}{14 TeV, 3000 fb$^{-1}$} \\
\cline{4-7}
  &  &  & $\Delta M \sim \mathcal{O}$(GeV) & & 400 & monojet & \\
  \hline
  
 \multirow{4}{*}{wino-higgsino} & \multirow{4}{*}{III} & \multirow{2}{*}{0.75} & $m_{\tilde{l}} > M_{\neu{3}}$ & 500 & 340 & \multirow{4}{*}{multi-lepton + $\met$} & \multirow{4}{*}{14 TeV, 3000 fb$^{-1}$} \\
  &  &  & $m_{\tilde{l}} < M_{\neu{3}}$ & 600 & 390 & & \\
 
\cline{3-6} 
  
  &  &  \multirow{2}{*}{1} & $m_{\tilde{l}} > M_{\neu{3}}$ & 550 & 250 & & \\
  &  &  & $m_{\tilde{l}} < M_{\neu{3}}$ & 600 & 590 & & \\
\hline  
\end{tabular}%
}
\end{center}
\caption{Summary of our $\gmu$ scan and subsequent collider analysis for non bino-like LSP. $\gmu$ allowed $M_{\neu{1}}$  at $2 \sigma$, have been tabulated for different conditions. In columns 5  the $\gmu$ allowed values presented are for $\tan\beta=50$. $\Delta M$ here stands for the mass-gap between $\chpm{1}$ and $\neu{1}$.}
\label{Conclusion2}
\end{table}

  In conclusion, if SUSY particles are culpable for the $\gmu$ excess, a vast region of the parameter space is within the exclusion reach of the proposed high luminosity LHC experiments. However, for higher masses of $M_{\neu{1}}$, $\gmu$ will be explained by a more compressed spectra and this suffers from a lack of $\met$ in the system, which is the most important ingredient to distinguish a SUSY signal from the SM background. A typical LHC exclusion reach for compressed spectra is predicted to be $\sim 325-375$ GeV for electroweakinos and $\sim 225-320$ GeV sleptons for mass-splittings $\sim 5-30$ GeV. For sleptons no definitive exclusion limit has been set so far for mass-splitting between 20 and 70 GeV.
  
   The signal sensitivities and mass reaches discussed thus far do not consider any systematic uncertainties. At high luminosity systematic uncertainties, due to upgraded detector designs and trigger conditions to counter high pile-up conditions, are expected. If we consider $10 \%$ systematic uncertainty on background estimation, the $95 \%$ CL exclusion reach of $\neu{2}$, for bino-like LSP, will reduce to 710 (850) GeV, 1025 (550) GeV and 1050 (825) GeV with $m_{\tilde{l}} < M_{\neu{2}}$ ($m_{\tilde{l}} > M_{\neu{2}}$) for Regions-I, II and III respectively. The corresponding reach for $\tilde{l}_L$ ($\tilde{l}_R$) will be 625 (525) GeV. We do not consider any systematic uncertainty on signal, since we have taken a conservative approach and used LO cross-sections of electroweakino pair productions only.
In addition a shape-based binned-likelihood analysis on single or multiple kinetic variables (e.g. $\met$, $M_{T_2}$, $p_T$ of the leading lepton) may improve the significances further.

Finally the next generation $\gmu$ experiment at FNAL should start running from 2016 and the improvement in experimental accuracy of $\gmu$ is expected to be four fold~\cite{FNAL:g-2}. The results from the aforementioned experiment will further constrain the SUSY parameter space. On the other hand the Fermi-LAT 10 years data on 40 dwarf galaxies and future $\gamma$-ray experiments like CTA are anticipated to improve the constraint on DM annihilation cross-section by another order of magnitude~\cite{CTA}. Similarly future direct detection experiments like XENON1T~\cite{Xenon1T} will improve DM-nucleon scattering cross-section by two orders of magnitude and perhaps find the LSP.


\section*{Acknowledgments}

 We thank Teruki Kamon, Louis Strigari, Joel Walker, Tao Han and Ranjan Laha for helpful discussions. We would also like to thank Azar Mustafayev for reading the manuscript. This work is supported in part by the DOE Grant No. DE-FG02-13ER42020 (B.D. and T.G.) and DE-FG02-12ER41808 (I.G. and Q.S.). This work used the Extreme Science
and Engineering Discovery Environment (XSEDE), which is supported by the National Science
Foundation grant number OCI-1053575. Open Science Grid~\cite{OSG} resources have also been used to generate large statistics background samples for collider studies. I.G. acknowledges support from the  Rustaveli
National Science Foundation  No. 03/79.


\section*{Appendix}
\appendix

\section{\label{DeltaM} Dependence of significance on $\Delta m_1$ and $\Delta m_2$}

The relative impact of mass gaps $\Delta m_1$ and $\Delta m_2$ are discussed in this Appendix. For this study we have set $M_2/\mu =0.75$, $\tan \beta = 10 $ for the BP $(M_{\neu{1}},m_{\tilde{l}}) = (250,275)$ GeV. The results are presented in Table~\ref{slltneu2Table}.

\begin{table}[!htp]
\begin{center}
\begin{tabular}{|c|c| c c| c |c c c|} 
\hline
\hline 
$M_2/\mu$ & $\tan \beta$ & ($M_{\neu{1}}$,$M_{\neu{2}}$) & ($m_{\tilde{l}}$,$m_{\tilde{\nu}_{l_L}}$) & $\Delta a_{\mu}$ & S & B & $\sigma$ \\  

          &   &    &    &  [$\times 10^{10}$]  &   &   &  \\
          &   & [GeV] & [GeV] &    &   &   &   \\      
\hline
\multirow{3}{*}{0.75} & \multirow{3}{*}{10} & \multirow{3}{*}{(250,300)} & (290,280) & 13.4 & 5109 & \multirow{3}{*}{2744} &  57.7  \\ 

   &   &   & (275,264) & 14.3 & 7577 &  & 74.6   \\ 

   &   &   & (260,248) & 15.3 & 7971 &  & 77.0   \\     
\hline
\end{tabular}
\end{center}
\caption{Mass spectrum and \gmu \, contribution of the benchmark points with $m_{\tilde{l}} < M_{\neu{2}}$ for different combination of $\Delta m_1$ and $\Delta m_2$ are presented. The corresponding signal and background rates along with significances expected at the LHC, at $\sqrt{s}=14$ TeV and for 3000 fb$^{-1}$ of integrated luminosity, are also shown. }
\label{slltneu2Table}
\end{table}

\section{\label{SS1}Search strategy for $3 \, l + 1 \, j + \met$ channel with $\Delta m \sim 50 $ GeV and $m_{\tilde{l}} > M_{\neu{2}} $ }

\begin{enumerate}
  \item $b$-veto, $\tau_h$-veto;
  \item Select exactly 1 jet, with $p_{T_j} > 30$ GeV and $|\eta_j| < 2.5$; 
  \item Select 3 isolated leptons~\footnote{Lepton isolation is parametrized by $I_{rel} < 0.15$, where $I_{rel}$ is the ratio of the scalar sum of the transverse momenta of hadrons and photons within $\Delta R = \sqrt{(\Delta \eta)^2+(\Delta\phi)^2}=0.4$ of the lepton, and the $p_T$ of the lepton.}, with $p_{T_l} > 7$ GeV and $|\eta_{l}| < 2.5$ ;       
  \item $Z$-veto (i.e. reject events with $70$ GeV $< M_{l^{\pm} l^{\mp}} <$  $110$ GeV);
  \item Select events with $12$ GeV $< min(M_{l^{\pm} l^{\mp}}) <$  $50$ GeV, where $min(M_{l^{\pm} l^{\mp}})$ is the minimum invariant mass of all possible  opposite sign same flavour (OSSF) lepton pairs;
  \item  $p_{T_{l_1}} < 50$ GeV, where $l_1$ is the leading lepton .
  \item  $\met > 50$ GeV; 
\end{enumerate}

  The efficiency of each cut on the signal and background are shown in Table~\ref{3lj} for the BP ($M_{\neu{1}},M_{\neu{2}}$)=(150,200) GeV with $M_2/\mu=2$ and $\tan \beta=10$. The signal consists of all possible combinations of electroweakino pair. The $VV \, + \,$ jets (where $V = W, \, Z$) background consists of up to 2-partons inclusive processes, while the $\ttbar \, + \,$ jets and $V \, + \,$ jets include up to 3-partons inclusive processes. Same method is used to generate the signal and backgrounds for all subsequent studies.

\begin{table}[!htp]
\begin{center}
\resizebox{\columnwidth}{!}{%
\begin{tabular}{|c|c|c|c|c|c|c|c|} 
\hline
\hline 
 Selection & Signal & $(t \rightarrow b  l  \nu) \,\bar{t}$ & $W \rightarrow l \nu$ & $Z \rightarrow l l$ &  $(W \rightarrow l \nu) \, W$ & $ (W \rightarrow l \nu) \, Z$ & $ZZ$ \\
 \hline
 
 before cuts                      & 1928  & 1.81$\times10^5$ & 3.72$\times10^7$ & 2.43$\times10^6$ & 4.82$\times10^4$ & 2.23$\times10^4$ & 2.16$\times10^4$ \\
 $b,\, \tau_h$-veto               & 1666  & 3.81$\times10^4$ & 3.43$\times10^7$ & 1.93$\times10^6$ & 4.02$\times10^4$ & 1.74$\times10^4$ & 1.56$\times10^4$ \\
 exactly 3$l$ and 1 jet           & 1.45 & 4.53 & -- & 4.22 & 0.02 & 73.7 & 4.82 \\
 $Z$-veto + $12 < M_{l^{\pm} l^{\mp}} < 50$  & 0.52 & 1.06  & -- & -- & 0.01 & 1.27 & 0.09 \\
  $p_{T_{l_1}} < 50$             & 0.31 & 0.60 & -- & -- & -- & 0.60 & 0.03 \\
  $\met > 50$    & 0.22 & 0.42 & -- & -- & -- & 0.39 & 0.01 \\
  
\hline
\end{tabular}%
}
\end{center}
\caption{[$3 \, l + 1 \, j + \met$ study] Summary of the effective cross-section (fb) for the signal and main sources of background at LHC14 for the BP ($M_{\neu{1}},M_{\neu{2}}$)=(150,200) GeV with $M_2/\mu=2$ and $\tan \beta=10$. ``--" indicates the background size is negligible.}
\label{3lj}
\end{table}


\section{\label{SS2}Search strategy for opposite-sign $2 \, l+ \geq 2 \, j + \met$ channel with $\Delta m \geq m_Z $ and $m_{\tilde{l}} > M_{\neu{2}}$ }

\begin{enumerate}
 \item $b$-veto;
 \item Select at least 2 jets, with $p_{T_j} > 30$ GeV and $|\eta_j| < 3$; 
 \item Select exactly 2 leptons, with $p_{T_l} > 20$ GeV and $|\eta_{l}| < 2.5$;
 \item Select events with at least a jet-pair satisfying $70$ GeV $< M_{jj} <$  $110$ GeV, where $M_{jj}$ is the invariant mass of any jet pair;
 \item Select events with OSSF lepton pair satisfying $70$ GeV $< M_{l^{\pm} l^{\mp}} <$  $110$ GeV;
 \item  $\met > 200$ GeV;
 \item Veto events with  40 GeV $< M_T < 150$ GeV, where the transverse mass, $M_T$, is formed from $\met$ and $p_{T_l}$ of the third remaining lepton and defined as $M_T = \sqrt{2 \met p_{T_l} (1-\cos(\Delta\phi_{l,\met}))}$.
 
\end{enumerate}

 The efficiency of each cut on the signal and background are shown in Table~\ref{2l2j} for the BP ($M_{\neu{1}},M_{\neu{2}}$)=(150,300) GeV with $M_2/\mu=2$ and $\tan \beta=50$. 

\begin{table}[!htp]
\begin{center}
\resizebox{\columnwidth}{!}{%
\begin{tabular}{|c|c|c|c|c|c|c|c|} 
\hline
\hline 
 Selection & Signal & $(t \rightarrow b  l \nu) \,\bar{t}$ & $W \rightarrow l \nu$ & $Z \rightarrow l l$ &  $(W \rightarrow l \nu) \, W$ & $ (W \rightarrow l \nu) \, Z$ & $ZZ$ \\
 \hline
 
 before cuts                      & 321  & 1.81$\times10^5$ & 3.72$\times10^7$ & 2.43$\times10^6$ & 4.82$\times10^4$ & 2.23$\times10^4$ & 2.16$\times10^4$ \\
 $b$-veto                         & 268  & 4.51$\times10^4$ & 3.70$\times10^7$ & 2.42$\times10^6$ & 4.81$\times10^4$ & 2.07$\times10^4$ & 1.86$\times10^4$ \\
 2$l$ and 2 jets                  & 6.69 & 3.80$\times10^3$ & 9.73 & 7.73$\times10^4$ & 852 & 178 & 522 \\
 $70 < M_{jj} < 110$              & 3.97 & 2.06$\times10^3$ & 2.43 & 2.82$\times10^4$ & 282 & 69.3 & 332 \\
 $70 < M_{l^{\pm} l^{\mp}} < 110$ & 1.93 & 296  & -- & 2.82$\times10^4$ & 37.7 & 35.9 & 324 \\
 $\met > 200$                     & 1.16 & 67.3 & -- & 3.16 & 10.3 & 9.53 & 5.10 \\
  $M_T$-veto                      & 0.44 & 6.82 & -- & 1.49 & 1.17 & 3.55 & 2.41 \\
  
\hline
\end{tabular}%
}
\end{center}
\caption{[OS $2 \, l + 2 \, j + \met$ study] Summary of the effective cross-section (fb) for the signal and main sources of background at LHC14 for the BP ($M_{\neu{1}},M_{\neu{2}}$)=(150,300) GeV with $M_2/\mu=2$ and $\tan \beta=50$.}
\label{2l2j}
\end{table}


\section{\label{SS3}Search strategy for $3 \, l + \met$ channel with $\Delta m \geq m_Z $ and and $m_{\tilde{l}} > M_{\neu{2}}$ }

\begin{enumerate}
  \item $b$-veto, $\tau_h$-veto;
  \item Select 3 isolated leptons, with $p_{T_{l_1}} > 20$ GeV, $p_{T_{l_2}} > 10$ GeV, $p_{T_{l_3}} > 10$ GeV and $|\eta_{l}| < 2.5$;  
  \item Require OSSF lepton pair;     
   \item Select events with OSSF lepton pair satisfying $70$ GeV $< M_{l^{\pm} l^{\mp}} <$  $110$ GeV;
   \item $\met > 200$ GeV.
   \item $\Delta \phi (\met, l_3) > 1$, where $l_3$ is the third remaining lepton; 
   \item Asymmetric $M_{T_2} > 250$ GeV, where asymmetric $M_{T_2}$ is computed out of the $\met$, the reconstructed $Z$-boson (OSSF lepton pair) as the visible particle on one chain and the third lepton on the other. $M_{T_2}$ algorithm of Ref.~\cite{MT2} has been adapted~\footnote{The algorithm of Ref.~\cite{MT2} is implemented by using the code made public by~\cite{MT2-math} and further validated against another publicly available code~\cite{MT2-perl} (See Ref.~\cite{Walker:MT2} for the details of this algorithm).}.

\end{enumerate}

 The cut-flow table for this analysis is presented in Table~\ref{3lWZ} for the BP\\ ($M_{\neu{1}},M_{\neu{2}}$)=(150,300) GeV with $M_2/\mu=2$ and $\tan \beta=50$. Asymmetric $M_{T_2}$ cut has not been used for the BP ($M_{\neu{1}},M_{\neu{2}},M_2/\mu,\tan \beta$)= (150 GeV, 250 GeV, 0.2, 10). 

\begin{table}[!htp]
\begin{center}
\resizebox{\columnwidth}{!}{%
\begin{tabular}{|c|c|c|c|c|c|c|c|} 
\hline
\hline 
 Selection & Signal & $(t \rightarrow b  l  \nu) \,\bar{t}$ & $W \rightarrow l \nu$ & $Z \rightarrow l l$ &  $(W \rightarrow l \nu) \, W$ & $ (W \rightarrow l \nu) \, Z$ & $ZZ$ \\
 \hline
 
 before cuts                      & 321  & 1.81$\times10^5$ & 3.72$\times10^7$ & 2.43$\times10^6$ & 4.82$\times10^4$ & 2.23$\times10^4$ & 2.16$\times10^4$ \\
 $b,\, \tau_h$-veto               & 222  & 3.81$\times10^4$ & 3.43$\times10^7$ & 1.93$\times10^6$ & 4.02$\times10^4$ & 1.74$\times10^4$ & 1.56$\times10^4$ \\
 exactly 3$l$                     & 2.89 & 40.9 & -- & 105 & 0.57 & 486 & 37.5 \\
 OSSF pair                        & 2.78 & 30.8 & -- & 104 & 0.44 & 483 & 37.2 \\
 $70 < M_{l^{\pm} l^{\mp}} < 110$ & 1.26 & 9.32  & -- & 98.2 & 0.12 & 467 & 36.0 \\
 $\met > 200$                     & 0.28 & 0.10 & -- & -- & -- & 9.26 & 0.13 \\
  $\Delta \phi (\met, l_3) > 1$   & 0.18 & 0.09 & -- & -- & -- & 1.10 & 0.01 \\
  Asymmetric $M_{T_2} > 250$   & 0.13 & 0.07 & -- & -- & -- & 0.02 & 0.002 \\
\hline
\end{tabular}%
}
\end{center}
\caption{[$3 \, l + \met$ study] Summary of the effective cross-section (fb) for the signal and main sources of background at LHC14 for the BP ($M_{\neu{1}},M_{\neu{2}}$)=(150,300) GeV with $M_2/\mu=2$ and $\tan \beta=50$.}
\label{3lWZ}
\end{table}
   

\section{\label{SS5}Search strategy for $4 \,l + \met$ channel with $\Delta m \geq m_Z $ and and $m_{\tilde{l}} > M_{\neu{2}}$ }

\begin{enumerate}
  \item $b$-veto;
  \item Select 4 isolated leptons, with $p_{T_{l_1}} > 20$ GeV, $p_{T_{l_{2,3,4}}} > 10$ GeV and $|\eta_{l}| < 2.5$;  
  \item Require two OSSF lepton pairs satisfying $70$ GeV $< M_{l^{\pm} l^{\mp}} <$  $110$ GeV;        
   \item $\met > 200$ GeV.

\end{enumerate}

 The cut-flow table for this analysis is presented in Table~\ref{4lZZ} for the BP \\ ($M_{\neu{1}},M_{\neu{2}}$)=(150,300) GeV with $M_2/\mu=2$ and $\tan \beta=50$. 

\begin{table}[!htp]
\begin{center}
\resizebox{\columnwidth}{!}{%
\begin{tabular}{|c|c|c|c|c|c|c|c|} 
\hline
\hline 
 Selection & Signal & $(t \rightarrow b  l  \nu) \,\bar{t}$ & $W \rightarrow l \nu$ & $Z \rightarrow l l$ &  $(W \rightarrow l \nu) \, W$ & $ (W \rightarrow l \nu) \, Z$ & $ZZ$ \\
 \hline
 
 before cuts                      & 321  & 1.81$\times10^5$ & 3.72$\times10^7$ & 2.43$\times10^6$ & 4.82$\times10^4$ & 2.23$\times10^4$ & 2.16$\times10^4$ \\
 $b$-veto               & 268  & 4.51$\times10^4$ & 3.70$\times10^7$ & 2.42$\times10^6$ & 4.81$\times10^4$ & 2.07$\times10^4$ & 1.86$\times10^4$ \\
 exactly 4 $l$          & 0.13 & 0.01 & -- & -- & -- & 0.02 & 25.6 \\
 2 OSSF pairs with      & \multirow{2}{*}{0.03} & \multirow{2}{*}{--} & \multirow{2}{*}{--} & \multirow{2}{*}{--} & \multirow{2}{*}{--} & \multirow{2}{*}{--} & \multirow{2}{*}{24.3} \\
 $70 < M_{l^{\pm} l^{\mp}} < 110$ &  &   &  &  &  &  &   \\
 $\met > 200$                     & 0.02 & -- & -- & -- & -- & -- & 0.02 \\  
  
\hline
\end{tabular}%
}
\end{center}
\caption{[$4 \, l + \met$ study] Summary of the effective cross-section (fb) for the signal and main sources of background at LHC14 for the BP  ($M_{\neu{1}},M_{\neu{2}}$)=(150,300) GeV with $M_2/\mu=2$ and $\tan \beta=50$.}
\label{4lZZ}
\end{table}


\section{\label{SS6}Search strategy for same-sign $2 \, l+  2/3 \, j + \met$ channel with $\Delta m \geq m_Z $ and $m_{\tilde{l}} > M_{\neu{2}}$ }

\begin{enumerate}
 \item $b$-veto;
 \item Select exactly 2 or 3 jets, with $p_{T_j} > 30$ GeV and $|\eta_j| < 3$; 
 \item Select exactly 2 isolated  same-sign leptons~\footnote{A tighter isolation criteria for the leptons of $I_{rel} < 0.10$  has been used for this study ( See Ref.~\cite{CMSSlep1} for details.) }, with $p_{T_l} > 20$ GeV and $|\eta_{l}| < 2.5$;
 \item Select events with at least one lepton satisfying  $ M_{T} >  110$ GeV, where $M_{T}$ is defined in Appendix~\ref{SS2};
 \item  $\met > 100$ GeV;
 
\end{enumerate}

 The efficiency of each cut on the signal and background are shown in Table~\ref{SSWh2l2j} for the BP ($M_{\neu{1}},M_{\neu{2}}$)=(150,300) GeV with $M_2/\mu=2$ and $\tan \beta=50$. 

\begin{table}[!htp]
\begin{center}
\resizebox{\columnwidth}{!}{%
\begin{tabular}{|c|c|c|c|c|c|c|c|} 
\hline
\hline 
 Selection & Signal & $(t \rightarrow b  l \nu) \,\bar{t}$ & $W \rightarrow l \nu$ & $Z \rightarrow l l$ &  $(W \rightarrow l \nu) \, W$ & $ (W \rightarrow l \nu) \, Z$ & $ZZ$ \\
 \hline
 
 before cuts                      & 321  & 1.81$\times10^5$ & 3.72$\times10^7$ & 2.43$\times10^6$ & 4.82$\times10^4$ & 2.23$\times10^4$ & 2.16$\times10^4$ \\
 $b$-veto                         & 268  & 4.51$\times10^4$ & 3.70$\times10^7$ & 2.42$\times10^6$ & 4.81$\times10^4$ & 2.07$\times10^4$ & 1.86$\times10^4$ \\
 SS 2$l$ and 2/3 jets                  & 0.13 & 0.97 & -- & -- & 4.78 & 44.7 & 0.31 \\
 $ M_{T} > 110$              & 0.12 & 0.10 & -- & -- & 2.38 & 15.6 & 0.07 \\ 
 $\met > 100$                     & 0.10 & 0.05 & -- & -- & 1.16 & 7.54 & 0.03 \\
 
  
\hline
\end{tabular}%
}
\end{center}
\caption{[SS $2 \, l + 2/3 \, j + \met$ study] Summary of the effective cross-section (fb) for the signal and main sources of background at LHC14 for the BP ($M_{\neu{1}},M_{\neu{2}}$)=(150,300) GeV with $M_2/\mu=2$ and $\tan \beta=50$.}
\label{SSWh2l2j}
\end{table}


\section{\label{SS4}Search strategy for $3 \, l + \met$ channel with  $m_{\tilde{l}} < M_{\neu{2}} $ }
\begin{enumerate}
  \item $b$-veto, $\tau_h$-veto;
  \item Select 3 isolated leptons, with $p_{T_{l_1}} > 20$ GeV, $p_{T_{l_2}} > 10$ GeV, $p_{T_{l_3}} > 10$ GeV and $|\eta_{l}| < 2.5$;  
  \item Require OSSF lepton pair;     
   \item $Z$-veto (i.e. reject events with $70$ GeV $< M_{l^{\pm} l^{\mp}} <$  $110$ GeV);   
   \item $\met > 200$ GeV;
   \item $p_{T_{l_1}} > 30-100$ GeV (optimized for each BP), where $l_1$ is the leading lepton.     
\end{enumerate}

  Table~\ref{3l} contains the signal and background yield of this analysis for the BP ($M_{\neu{1}},M_{\neu{2}},m_{\tilde{l}}$)=(150,400,175) GeV with $M_2/\mu=2$ and $\tan \beta=10$.

\begin{table}[!htp]
\begin{center}
\resizebox{\columnwidth}{!}{%
\begin{tabular}{|c|c|c|c|c|c|c|c|} 
\hline
\hline 
 Selection & Signal & $(t \rightarrow b  l  \nu) \,\bar{t}$ & $W \rightarrow l \nu$ & $Z \rightarrow l l$ &  $(W \rightarrow l \nu) \, W$ & $ (W \rightarrow l \nu) \, Z$ & $ZZ$ \\
 \hline
 
 before cuts                      & 102  & 1.81$\times10^5$ & 3.72$\times10^7$ & 2.43$\times10^6$ & 4.82$\times10^4$ & 2.23$\times10^4$ & 2.16$\times10^4$ \\
 $b,\, \tau_h$-veto               & 76.6  & 3.81$\times10^4$ & 3.43$\times10^7$ & 1.93$\times10^6$ & 4.02$\times10^4$ & 1.74$\times10^4$ & 1.56$\times10^4$ \\
 exactly 3$l$                     & 4.03 & 40.9 & -- & 105 & 0.57 & 486 & 37.5 \\
 OSSF pair                        & 3.98 & 30.8 & -- & 104 & 0.44 & 483 & 37.2 \\
 $Z$-veto                         & 2.61 & 21.7 & -- & 7.39 & 0.32 & 15.5 & 1.27 \\
 $\met > 200$                     & 1.23 & 0.40 & -- & -- & 0.01 & 0.53 & 0.03 \\
  $p_{T_{l_1}} > 100$             & 1.23 & 0.15 & -- & -- & -- & 0.20 & 0.01 \\
  
\hline
\end{tabular}%
}
\end{center}
\caption{[$3 \, l  + \met$ study for $m_{\tilde{l}} < M_{\neu{2}}$] Summary of the effective cross-section (fb) for the signal and main sources of background at LHC14 for the BP ($M_{\neu{1}},M_{\neu{2}},m_{\tilde{l}}$)=(150,400,175) GeV with $M_2/\mu=2$ and $\tan \beta=10$. }
\label{3l}
\end{table}


\newpage


\end{document}